\documentclass[11pt]{article}

\usepackage{euscript}
\usepackage{amssymb}
\usepackage{amsfonts}
\usepackage{amsbsy}
\usepackage{amsmath}
\usepackage{epsfig}
\usepackage{amsthm}
\usepackage{amscd}
\usepackage{amstext}
\usepackage{verbatim}

\usepackage[pdftex]{hyperref}

\def\ben{\begin{equation}}
\def\een{\end{equation}}
\def\half{{\textstyle{1\over2}}}

\let\a=\alpha  \let\g=\gamma  
   \let\k=\kappa
     
 \let\t=\tau

\let\w=\omega \let\G=\Gamma

\let\pa=\partial
\def\be{\begin{equation}}
\def\ee{\end{equation}}
\def\ba{\begin{array}}
\def\ea{\end{array}}

\def\dalemb#1#2{{\vbox{\hrule height .#2pt
       \hbox{\vrule width.#2pt height#1pt \kern#1pt
               \vrule width.#2pt}
       \hrule height.#2pt}}}

\newcommand{\bea}{\begin{eqnarray}}
\newcommand{\eea}{\end{eqnarray}}
\newcommand{\tr}{{\rm tr} }

\def\R{{{\Bbb R}}}

\def\ocal{{\mathcal{O}}}

\newcommand{\gammat}{\Gamma^{\underline{t}}}
\numberwithin{equation}{section}

\begin{document}

\ \\
\vspace{1cm}

\begin{center}

{ \LARGE {\bf Cooper pairing near charged black holes}}

\vspace{1.2cm}

Thomas Hartman and Sean A. Hartnoll

\vspace{0.9cm}

{\it Department of Physics, Harvard University,
\\
Cambridge, MA 02138, USA \\}

\vspace{0.5cm}

{\tt  hartman@physics.harvard.edu, hartnoll@physics.harvard.edu} \\

\vspace{1.3cm}

\end{center}

\begin{abstract}

We show that a quartic contact interaction between charged fermions can lead to Cooper pairing
and a superconducting instability in the background of a charged asymptotically Anti-de Sitter
black hole. For a massless fermion we obtain the zero mode analytically and compute the dependence of the critical temperature $T_c$ on the charge of the fermion.
The instability we find occurs at charges above a critical value, where
the fermion dispersion relation near the Fermi surface is linear.
The critical temperature goes to zero as the marginal Fermi liquid is approached, together with the density of states at the Fermi surface. Besides the charge, the critical temperature is controlled by a four point function of a fermionic operator in the dual strongly coupled field theory.

\end{abstract}

\pagebreak
\setcounter{page}{1}

\section{Introduction}

The first successful microscopic description of superconductivity, BCS theory \cite{BCS}, describes spontaneous symmetry breaking due to
a charged fermion bilinear condensate. In the original theory the fermion pairing is driven by an attractive exchange of low energy phonons. More generally, the essential feature is a marginally relevant four point interaction between excitations about a Fermi surface \cite{Polchinski:1992ed}. Whether this interaction is generated by phonons or otherwise is not crucial. The key fact is rather that in BCS-like theories, superconductivity emerges from a conventional free Fermi liquid fixed point. An important challenge facing condensed matter theory is to characterise the onset of superconductivity from non-Fermi liquid states of matter, such as the `strange metal' phases of high temperature superconductors, e.g. \cite{hussey1, hussey2}.

Recent developments have shown that charged or rotating black holes can carry Fermi surfaces \cite{Lee:2008xf, Liu:2009dm, Cubrovic:2009ye, Faulkner:2009wj, Hartnoll:2009ns,Hartman:2009qu,Faulkner:2010tq}. We focus on the charged case in 3+1 dimensional asymptotically Anti-de Sitter (AdS) spacetime in what follows; it seems clear that many features of the computation will go through in the rotating case also and, with nonzero fermion mass, in asymptotically flat space. Near to the black hole, charged fermions experience a background chemical potential. If the charge of the fermions is sufficiently large compared to their mass, then at low Hawking temperatures they will build up a Fermi surface. The essential physics is the same as that for free fermions in flat space with a chemical potential. Computations are complicated by the fact that the background spacetime and electrostatic potential are nontrivial and vary on scales of order the Compton wavelength of the fermions. In general the Dirac equation cannot be solved in closed form on the whole spacetime.

It is natural to ask whether black hole Fermi surfaces can have BCS instabilities towards superconductivity. In this paper we will add a four fermion contact interaction between the charged fermions and compute the quadratic term of the one loop effective action for Cooper pairs. We will show that under certain circumstances the quadratic action has negative modes, indicating a superconducting instability. The effective action is not local in general on the curved spacetime background; this complicates finding e.g. the zero temperature gap. However, we have been able to obtain an analytic formula for the critical temperature $T_c$. Furthermore, for massless fermions we have found the Fermi surface zero mode analytically, allowing explicit results without heavy duty numerical work. The critical temperature is
\be
T_c \propto \mu \, e^{-M_F^2 L^2/N_\text{eff.}} \,,
\ee
where $\mu$ is the chemical potential provided by the charged black hole at the AdS boundary, $M_F$ is the energy scale of the four fermion interaction, $L$ is the AdS radius, and $N_\text{eff.}$ is the effective density of states at the Fermi surface. From the perspective of the dual field theory, the dimensionless quantity $M_F L$ determines the magnitude of a four point fermion correlator. Figure \ref{fig:Neff} shows the fermion charge dependence of $N_\text{eff.}$, which is  given by
\be
N_\text{eff.} \sim k_F v_F \int\sqrt{-g}(\psi^{0\dagger}\psi^0)^2 \,,
\ee
where $k_F$ is the Fermi momentum, $v_F$ is the Fermi velocity, and $\psi^0$ is the fermion zero mode in the black hole spacetime. The precise formula is given in (\ref{eq:n}) below.

For asymptotically AdS charged back holes the bulk (free) Fermi surface admits a dual interpretation, via the applied holographic correspondence \cite{Maldacena:1997re, Hartnoll:2009sz, Herzog:2009xv, McGreevy:2009xe, Hartnoll:2009qx}, as a strongly interacting (non-)Fermi liquid in 2+1 dimensions \cite{Lee:2008xf, Liu:2009dm, Cubrovic:2009ye, Faulkner:2009wj}. It was understood in \cite{Faulkner:2009wj} that for fermions with a relatively low charge compared to their mass, the dispersion relation of fermion zero modes near the Fermi surface had a non-Fermi liquid form. Furthermore, these modes were broad resonances in the spectral density, rather than sharp quasiparticle peaks, as is indeed observed in strange metals \cite{ARPES}. The non-Fermi dispersion was shown to lead to, for instance, deviations from the venerable Lifshitz-Kosevich formula for quantum oscillations \cite{Denef:2009yy, Hartnoll:2009kk}. The black hole BCS instability we present below therefore has the potential to dually describe the non-BCS emergence of superconductivity from a strongly interacting non-Fermi liquid. Unfortunately, perhaps, we will find that the superconducting instability only occurs at larger values of the fermion charge, where the dispersion relation is linear (i.e. Fermi liquid like). This can be traced directly to the vanishing of the density of states at the Fermi surface in the non-Fermi liquid cases. It may be possible to evade this conclusion via alternate bulk pairing mechanisms with long range interactions.

Recent works have considered instabilities of charged scalar fields in charged black hole backgrounds and the corresponding spontaneous symmetry breaking at low temperatures \cite{Gubser:2008px, Hartnoll:2008vx, Hartnoll:2008kx}. If the charge of the boson is sufficiently large compared to its mass \cite{Denef:2009tp} it will condense, again in strong analogy to the behaviour of charged bosons in flat space with a chemical potential. As with the Cooper pairing instability we have just outlined, the dual interpretation is of superconductivity emerging from a strongly interacting non-Fermi liquid. One difference is that the boson condensation is classical in the black hole background whereas for fermions the effect requires a one loop computation, with an ensuing nonlocal (bulk) Landau-Ginzburg action. In the Cooper pairing case of interest here, the superconducting order parameter is directly related to a fermionic operator in the dual field theory. This may be phenomenologically useful and motivates fermion spectroscopy (`ARPES') computations in the superconducting state along the lines of \cite{Faulkner:2009am}, in which the fermion is chosen to couple in a natural way to the bosonic condensate.

\section{Selfinteracting Dirac fermion}

We consider a charged Dirac fermion with quadratic action
\be\label{eq:dirac}
S_\text{Dirac} = \int d^4x \sqrt{-g}\ i\left(\bar\psi \Gamma^\mu D_\mu \psi - m \bar\psi \psi\right) \ ,
\ee
where
\be\label{covder}
D_\mu = \partial_\mu + {1\over 4}\omega_{ab\mu}\Gamma^{ab} - iq A_\mu \,,
\ee
$\bar\psi = \psi^\dagger \Gamma^{\underline{t}}$, $\Gamma^{ab} = \Gamma^{[a}\Gamma^{b]}$,  $\omega_{ab\mu}$ is the spin connection, and $\Gamma^\mu \Gamma^\nu + \Gamma^\nu \Gamma^\mu  = 2g^{\mu\nu}$, with a mostly plus metric. We denote bulk spacetime indices by $\mu, \nu, \dots$, abstract tangent space indices by $a,b,\cdots$, and specific tangent space indices by underlines as in $\Gamma^{\underline{t}}$. Eventually we will take the background to be a charged black hole in AdS, but for now the metric and gauge field are general.

The BCS mechanism requires an attractive force between like-charge particles.  We therefore add the simple contact interaction
\be\label{intera}
S_\text{int} = {1 \over M_F^2} \int d^4x \sqrt{-g} (\bar\psi_c \Gamma^5 \psi)(\bar\psi \Gamma^5 \psi_c) \,,
\ee
where $M_F$ is the mass scale of the interaction, $\Gamma^5 = i\Gamma^{\underline{0}}\Gamma^{\underline{1}} \Gamma^{\underline{2}} \Gamma^{\underline{3}}$,  and $\psi_c$ is the charge conjugate fermion
\be
\psi_c = C \bar\psi^T  \ , \quad C^{-1}\Gamma^aC = - (\Gamma^a)^T \ .
\ee
The interaction (\ref{intera}), which also appears in color superconductivity \cite{Alford:2007xm}, is the relativistic generalization of $s$-wave BCS theory: it couples time-reversed, opposite spin states \cite{Bertrand,BertrandGovaerts}. A similar interaction was considered in a closely related context in \cite{Faulkner:2009am}. However, this choice is not unique.  Besides choosing a more general contact term, the pairing mechanism could arise from exchange of scalar particles, the attractive channel in a nonabelian gauge theory, or perhaps graviton exchange. An attractive interaction per se is not sufficient to generate superconductivity, but should be in a `Cooper channel'. The contact interaction (\ref{intera}) is simpler than an exchange interaction, and can be considered a toy model for these other possibilities which may be more natural from the standpoint of string theory on AdS.

\section{Effective action for the condensate}

Mimicking the standard procedure in BCS theory, we can now perform a Hubbard-Stratanovich decoupling to make the action quadratic in spinors. As usual, there is a choice of channels to decouple. Given that we are anticipating a superconducting instability of the Fermi surface, we choose to decouple in the Cooper channel. Thus we introduce a charged scalar $\Delta$ and write the Lagrangian as
\be\label{hsint}
\mathcal{L}_\text{int} = \bar\psi_c \Gamma^5 \psi \Delta + \bar\psi\Gamma^5 \psi_c \Delta^* - M_F^2| \Delta|^2 \ .
\ee
Recall that the fermions anticommute.
The equation of motion for $\Delta$ sets
\be
\Delta = {1 \over  M_F^2}\bar\psi \Gamma^5 \psi_c \ , \quad \Delta^* =  {1 \over  M_F^2} \bar\psi_c \Gamma^5 \psi \ ,
\ee
and we recover the original action.

Now consider the Coleman-Weinberg effective action for $\Delta$, to look for possible instabilties. Specifically, we will compute the one loop mass term generated for $\Delta$ upon integrating out the fermions. The effective action at quadratic order is
\bea\label{eq:eff1}
\lefteqn{S_\text{eff}^{(2)}[\Delta] =  M_F^2 \int d^4x \sqrt{g}|\Delta(x)|^2 } \nonumber \\
& & \, -2\int d^4x d^4x' \sqrt{g(x)}\sqrt{g(x')}\Delta(x)\Delta^*(x') \tr \, G^T(x,x')C\Gamma^5G(x,x')C\Gamma^5 \,.
\eea
Here $G(x,x')$ is the Euclidean Green's function for the Dirac operator in the gauge field and spacetime background, $G(x,x') = -\langle \psi(x)\bar\psi(x')\rangle$. $G^T(x,x')$ is the transpose of the Green's function in spin indices, i.e. $G^T_{st}(x,x') = G_{ts}(x,x')$. To derive this expression we used $C = C^\dagger = - C^T$, $\G^{5\, T} = \G^5$ and $[C,\G^5] = 0$. See the representation of the gamma matrices in equation (\ref{gammas}) below. Note also that the interaction term in the Euclidean action is minus that in the Lorentzian action. We use Lorentzian gamma matrices throughout.

We now choose coordinates $\{u,\t,\vec x\}$, with $\tau$ Euclidean time, and assume the spacetime is translationally invariant along $\{\tau,\vec{x}\}$. In AdS, the radial coordinate is $u$ and the boundary directions are $\{\tau, \vec{x}\}$.  Thus we can Fourier transform
\be\label{eq:Gfourier}
G(x,x') = T \sum_n \int \frac{d^2k}{(2\pi)^2} G(u,u',i \w_n,k) e^{- i \w_n(\t-\t') + i \vec k \cdot (\vec x- \vec x')} \,,
\ee
where the fermionic Matsubara frequencies at temperature $T$ are
\be
\w_n = \pi T ( 2 n + 1) \,.
\ee
We furthermore restrict to configurations in which the condensate $\Delta = \Delta(u)$ only depends on the radial direction.
The effective action (\ref{eq:eff1}) becomes
\be\label{eq:effectiveaction}
S_\text{eff}^{(2)}[\Delta] =  M_F^2 \frac{V_2}{T} \int du \sqrt{g}|\Delta(u)|^2  + {V_2\over T} \int du du' \sqrt{g(u)g(u')}\Delta(u)  \Delta^*(u') F(u,u') \,,
\ee
where $V_2$ is the boundary spatial volume and
\be\label{eq:definef}
F(u,u') = -2T  \sum_n \int \frac{d^2k}{(2\pi)^2}\tr \, G^T(u,u',i\w_n, \vec k)C\Gamma^5 G(u,u',-i \w_n,-\vec k)C\Gamma^5 \ .
\ee

The next step is to relate the Euclidean Green's functions appearing in (\ref{eq:definef}) to real time Green's functions.
This is a little subtle, although the bottom line is that the boundary conditions at the black hole horizon mimic the usual effects of finite temperature field theory. In particular, as emphasized in \cite{Denef:2009yy, Denef:2009kn}, eigenfunctions and eigenvalues, and hence Green's functions, are not analytic functions of $i \w_n$. This is because regularity at the Euclidean `horizon' $u=u_+$ typically requires behaviour of the form
\be\label{finitethorizon}
\psi \sim (u-u_+)^{|\w_n|/(4 \pi T)} \,.
\ee
The positive and negative thermal frequencies must therefore be analytically continued separately. Analytically continuing, by setting $i \w_n \to z$,
the Euclidean Green's function from the upper imaginary frequency axis yields the retarded Green's function $G^R(z,\vec k)$, with poles
in the lower half frequency plane. Analytic continuation from the lower imaginary frequency axis gives the advanced Green's function, $G^A(z,\vec k)$.
This relation between Euclidean, retarded and advanced Green's functions is a general statement that is particularly transparent in the black hole context, as we recall in an appendix.

The sum over Matsubara frequencies can therefore be rewritten as a contour integral
\bea
\lefteqn{T \sum_n \tr \, G^T(u,u',i\w_n, \vec k)C\Gamma^5 G(u,u',-i \w_n,-\vec k)C\Gamma^5} \nonumber \\
&& = {i\over 4\pi}
\int_{\mathcal{C}} dz \, \tr \, G^T(u,u', z , \vec k)C\Gamma^5 G(u,u',- z,-\vec k)C\Gamma^5\tanh\left(z\over 2T\right)  \ ,\label{eq:gtanh}
\eea
where the contour $C$ has a segment in the upper half plane and a segment in the lower half plane, each going clockwise around the poles of $\tanh$. The analytically continued function $G$ has a branch cut on the real $z$ axis.  In the upper half plane, schematically,
\be\label{continueg}
G(z)G(-z) = G^R(z) G^A(-z) \ ,
\ee
where $G^{R,A}$ are the retarded and advanced Green's functions. This product is analytic in the upper half plane.  In the lower half plane $G^A$ and $G^R$ are exchanged.  On the real axis the correlators are related by
\be\label{eq:realaxis}
G^A(u,u', \Omega, \vec{k}) =  \gammat G^R(u',u,\Omega, \vec{k})^\dagger \gammat \ ,
\ee
where the transpose in $G^{R \dagger}$ acts on spin indices. This result follows easily from the definition of the various Green's functions, see the appendix. Deforming the contours in (\ref{eq:gtanh}) onto the real axis then gives\be
F(u,u') =- i \int \frac{d^2k}{(2\pi)^2} \int_{-\infty}^\infty \frac{d \Omega}{\pi} \, \tanh \frac{\Omega}{2T}
 \tr \, \gammat G^R(u',u,\Omega, \vec k)^* \gammat C\Gamma^5 G^R(u,u',-\Omega,-\vec k)C\Gamma^5
  \,.\label{eq:eff3}
\ee
Our objective now is to evaluate these integrals.



\section{The charged AdS black hole}

At this point we will specialize to a planar, charged, asymptotically AdS black hole background. This is a solution to Einstein-Maxwell theory
\be\label{eq:einsteinmaxwell}
S_{\{g,A\}} = \int d^{4}x \sqrt{-g} \left(  \frac{1}{2 \kappa^2} \left(R + \frac{6}{L^2} \right)   - \frac{1}{4 g^2}  F^2 \right)\,.
\ee
The black hole background is given by
\be
ds^2 = \frac{L^2}{u^2} \left(- f(u) dt^2 + \frac{du^2}{f(u)} + dx^2 + dy^2 \right) \,, \qquad A = \Phi(u) dt \,,
\ee
with
\be
f = 1 - \left(1 + \frac{u_+^2 \mu^2}{\gamma^2} \right) \left(\frac{u}{u_+}\right)^3 +
\frac{u_+^2 \mu^2}{\gamma^2} \left(\frac{u}{u_+}\right)^{4} \,.
\ee
The horizon is at $u=u_+$ and the conformal boundary is $u=0$.
The chemical potential $\mu$ of the dual CFT is the boundary value of the
Maxwell potential
\be
\Phi = \mu \left(1 - \frac{u}{u_+}\right) \,.
\ee
We also introduced the ratio of electric and gravitational couplings
\be\label{eq:gamma}
\gamma^2 = \frac{2 g^2 L^2}{\kappa^2} \,.
\ee
In terms of the above quantities, the Hawking temperature of the black hole (and temperature of the dual field theory) is
\be\label{eq:T}
T = \frac{|f'(u_+)|}{4\pi} = \frac{1}{4 \pi u_+} \left(3 -  \frac{u_+^2 \mu^2}{\gamma^2} \right) \,.
\ee
The nonzero components of the spin connection are
\be\label{spincon}
\w^{ab}_t = \delta_t^{[a} \delta_u^{b]} u^2 \left( \frac{f}{u^2} \right)' \,, \qquad
\w^{ab}_i = - \delta_i^{[a} \delta_u^{b]} \frac{2 \sqrt{f}}{u} \,.
\ee
Finally, we adopt the following gamma matrix conventions of \cite{Faulkner:2009wj, Faulkner:2009am}. These are useful for simplifying the Dirac equation once rotational invariance has been used to consider momentum in the $x$ direction without loss of generality. Thus
\be\label{gammas}
\Gamma^{\underline{t}} = \left(\!
                           \begin{array}{cc}
                             i \sigma^1 & 0 \\
                             0 & i \sigma^1 \\
                           \end{array}
                        \! \right)
, \quad
\Gamma^{\underline{u}} = \left(\!
                           \begin{array}{cc}
                             -\sigma^3 & 0 \\
                             0 & -\sigma^3 \\
                           \end{array}
                         \!\right)
, \quad
\Gamma^{\underline{x}} = \left(\!
                           \begin{array}{cc}
                             -\sigma^2 & 0 \\
                             0 & \sigma^2 \\
                           \end{array}
                         \!\right)
, \quad
\Gamma^{\underline{y}} = \left(\!
                           \begin{array}{cc}
                             0 & \sigma^2 \\
                             \sigma^2 & 0 \\
                           \end{array}
                         \!\right) \,.
\ee
In this representation the charge conjugation matrix is $C = \Gamma^{\underline{t}}\Gamma^{\underline{u}}$. This representation is not quite the same as in \cite{Faulkner:2009wj, Faulkner:2009am} because our radial coordinate $u$ is the inverse of their radial coordinate.

\section{The retarded bulk Green's function and $T_c$}
\label{sec:retarded}

In order to compute the effective action of the condensate (\ref{eq:effectiveaction}) in the black hole background, we need the retarded Green's function.  It is the unique solution of
\be
(\Gamma^\mu D_\mu - m)G^R(x,x') = {1\over \sqrt{-g}}i\delta^{(4)}(x,x') \,,
\ee
subject to certain boundary conditions discussed below.  Transforming into momentum space except in the radial direction,
\be\label{kgreen}
D(\Omega,k)G^R(u, u', \Omega, k) = {1\over \sqrt{-g}}i\delta(u,u') \,,
\ee
where $D(\Omega,k)$ is the radial Dirac operator (including the mass term). The Green's function equation is solved in the usual manner by multiplying together solutions of the homogeneous equation with a discontinuity across the delta function.  For a mode with momentum in the $x$ direction,
\be
\psi = e^{-i\Omega t + i k x}\psi_\text{radial}(u) \ ,
\ee
the Dirac equation is
\be\label{radialdiracmom}
D(\Omega,k)\psi_\text{radial}(u) = 0 \ .
\ee
This equation is written out explicitly in (\ref{eq:sepdirac}) but here we need only some general properties.  Writing the wavefunction as
\be\label{projsplit}
\psi_\text{radial}(u) = \left(\!
                          \begin{array}{c}
                            \psi_1 \\
                            \psi_2 \\
                          \end{array}
                        \!\right) \ ,
\ee
the two-component spinors $\psi_1$ and $\psi_2$ decouple.  The gamma matrices (\ref{gammas}) were chosen as in \cite{Faulkner:2009wj, Faulkner:2009am} to make this decoupling manifest.

The retarded Green's function satisfies ingoing boundary conditions at the horizon.  Near the AdS boundary ($u=0$),
the leading solution of the wave equation is
\be
\psi_\alpha \sim
a_\alpha u^{3/2-mL} \left(\!
               \begin{array}{c}
                 1 \\
                 0 \\
               \end{array}
            \! \right)
 + b_\alpha u^{3/2+mL}
        \left(\!
               \begin{array}{c}
                 0 \\
                 1 \\
               \end{array}
            \! \right) \,.
\ee
The two spinors $(1,0)$ and $(0,1)$ are eigenspinors of $\Gamma^{\underline{u}}$ with opposite eigenvalues, implying that $a_\alpha$ and $b_\alpha$ are canonically conjugate (in a radial Hamiltonian slicing). See e.g. \cite{Iqbal:2009fd}. Therefore a boundary condition must be imposed on one, with the other allowed to fluctuate.  When $mL>\half$, we must pick the fluctuating piece to be the normalizable mode proportional to $(0,1)$.  More generally, for any $m$ we will choose to impose the boundary condition $a_\alpha =0$ on the fluctuating mode.\footnote{For $mL<\half$, there is another possible choice $b_\alpha = 0$ discussed in e.g. \cite{Faulkner:2009wj}, and for $m=0$ there is a continuous choice of boundary conditions discussed in e.g. \cite{Porrati:2009dy,Rattazzi:2009ux}.  These more general quantization choices will not be considered here.}

With these boundary conditions, the solution of (\ref{kgreen}) is constructed as follows. For each $\alpha=1,2$, take $\psi^\text{in}_\alpha$ to be the solution ingoing at the horizon, and $\psi^\text{bdy}_\alpha$ to be the solution with $a_\alpha = 0$ near the boundary.  Then
\be
G^R = G^R_1 \oplus G^R_2 \,,
\ee
where
\be\label{eq:explicit}
G^R_\alpha ={i\over W(\psi^\text{in}_\a,\psi^\text{bdy}_\a) } \times
\begin{cases}
\psi^\text{in}_\alpha(u)\widetilde\psi^\text{bdy}_\a(u') & u>u'\\
\psi^\text{bdy}_\alpha(u)\widetilde\psi^\text{in}_\a(u') & u<u'
\end{cases} \,,
\ee
with $\widetilde\psi_\a \equiv i \psi_\a^T \sigma^1$.
The Wronskian $W$ is a constant related to the conserved charge current:
\be
W(\chi,\psi) \equiv -\frac{1}{2} \sqrt{-g} \sqrt{g^{uu}}  \left(\widetilde\psi \sigma^3 \chi - \widetilde\chi \sigma^3 \psi\right) \ .
\ee
At general frequency and momentum $\{\Omega, k\}$, the solution satisfying ingoing boundary conditions at the horizon will not satisfy the asymptotic boundary condition, so generically $\psi^\text{in}_\a \neq \psi^\text{bdy}_\a$. We will normalize the solutions so that near the asymptotic boundary $u \to 0$ we have
\be\label{wavenormalization}
\psi^\text{bdy}_\a = u^{3/2+m L} \left(\!
               \begin{array}{c}
                 0 \\
                 1 \\
               \end{array}
            \! \right)
  + \cdots \ , \quad
\psi^\text{in}_\a = {1\over \mathcal{G}_\alpha}
 u^{3/2-m L} \left(\!
               \begin{array}{c}
                 1 \\
                 0 \\
               \end{array}
            \! \right)
 + u^{3/2+m L}
        \left(\!
               \begin{array}{c}
                 0 \\
                 1 \\
               \end{array}
            \! \right)  + \cdots \,.
\ee
We have introduced the quantity ${{\mathcal G}_\alpha(\Omega, k)}$, which is the retarded Green's function of the boundary field theory,
similar to the original discussion (for bosons) in \cite{Son:2002sd}. The Dirac equation (\ref{radialdiracmom}) is a real equation for $\psi_\a$, so the boundary condition implies $\psi^\text{bdy}$ is real.  By evaluating the $u$-independent Wronskian near the asymptotic boundary we obtain
\be\label{eq:wronskivalue}
W(\psi^\text{in}_\a,\psi^\text{bdy}_\a) = \frac{- iL^3}{{{\mathcal G}_\alpha(\Omega, k)}} \,.
\ee
We can see immediately that poles of the boundary and bulk Green's functions occur at the (in general complex) quasinormal frequencies of the black hole background, at which the mode satisfies both the horizon and asymptotic boundary conditions (see e.g. \cite{Ching:1995tj}).

It follows from the gamma matrices (\ref{gammas}) that the upper and lower spinor projections $\psi_{1,2}$ are related by $k \to -k$. Henceforth we drop the subscript and rewrite all quantities in terms of the first projection, denoting
\be
\psi \equiv \psi_1 \ , \quad \mathcal{G} \equiv \mathcal{G}_1 \ .
\ee
We can now rewrite (\ref{eq:eff3}) using (\ref{eq:explicit}), (\ref{eq:wronskivalue}), (\ref{gammas}),  (\ref{projsplit}), and rotational invariance, as
\bea\label{eq:eff4}
F(u,u') &=&  {i\over L^6} \int_{-\infty}^\infty \frac{|k|dk}{2\pi}\int_{-\infty}^\infty \frac{d \Omega}{\pi}\, \tanh \frac{\Omega}{2T} \mathcal{G}(\Omega, k)^* \mathcal{G}(-\Omega, k) \, \times \\
& & \begin{cases}
\psi^\text{bdy}(\Omega, u')^\dagger\psi^\text{bdy}(-\Omega, u')
\psi^\text{in}(\Omega, u)^\dagger\psi^\text{in}(-\Omega, u)  & u>u'\\
\psi^\text{in}(\Omega, u')^\dagger\psi^\text{in}(-\Omega, u')
\psi^\text{bdy}(\Omega, u)^\dagger\psi^\text{bdy}(-\Omega, u)  & u<u'\\
\end{cases}\ .\nonumber
\eea
All wavefunctions in this expression have momentum $k\hat{x}$.  We have used the identity
\be\label{grgrid}
-\gammat G_R(u',u,\Omega,\vec{k})^*\gammat =G_R(u,u',\Omega,\vec{k})^\dagger \ ,
\ee
or equivalently, using (\ref{eq:realaxis}), $G_A(u,u', \Omega, \vec{k}) = -G_R(u,u',\Omega,\vec{k})^*$. This last statement can be seen directly from (\ref{eq:explicit}) and the fact that $\psi^\text{bdy}$ is real and $\psi^\text{in *} = \psi^\text{out}$, see the appendix. Note the modulus sign on $|k|$ and that the integral is over both positive and negative $k$. Positive and negative momenta should be thought of as corresponding to the $ \mathcal{G}_1$ and $ \mathcal{G}_2$ components of the full Green's function, respectively.

Our first objective is to perform the frequency and momentum integrals in (\ref{eq:eff4}) (with some regulator). In principle all the quantities appearing in this expression could be obtained numerically. However, we can do better at low temperatures $T \ll \mu$. It was observed numerically in \cite{Lee:2008xf, Liu:2009dm, Cubrovic:2009ye}, and analytically in \cite{Faulkner:2009wj}, that at these low temperatures poles in the retarded Green's function can move close to the origin of the complex frequency plane. Specifically, the pole location $\w_\star(k) \sim 0$ as $k \sim k_F$.
This is of course the signature of a Fermi surface in the bulk geometry, as we might expect for charged fermions in a background electrostatic potential.
Following the experience of BCS theory, which we are essentially replicating in a curved spacetime background, one might expect that the crucial pairing physics occurs close to the Fermi surface.

Close to the Fermi momentum, we will recall in the next section that \cite{Faulkner:2009wj}
\be\label{eq:GF}
{\mathcal G}(\Omega,  k) = \frac{h_1}{k_\perp - \Omega/v_F + T^{2 \nu} {\mathcal F}_{\nu}(\frac{\Omega}{T})} + \cdots \,.
\ee
Here $k_\perp=k-k_F$ is the perpendicular distance of the momentum from the Fermi surface, $h_1$ and $v_F$ are real constants, and $\nu$ is a zero temperature critical exponent to be described below. The function $T^{2\nu}{\mathcal F}_{\nu}$ is the near horizon Green's function, that will also be characterised in the following section and in an appendix. At low temperatures, $\Omega/T \to \infty$,
\be\label{eq:lowTlimit}
T^{2 \nu} {\mathcal F}_{\nu}{\textstyle \left(\frac{\Omega}{T}\right)} = h_2 e^{i \theta - i \pi \nu} \Omega^{2\nu} + \cdots \,,
\ee
where $h_2$ is positive and the phase $\theta$ is such that poles of (\ref{eq:GF}) are in the lower half complex frequency plane.

We expect the dominant contribution to be due to the singular locus of the boundary Green's function (\ref{eq:GF}). Thus we restrict consideration to near the Fermi surface\footnote{As well as near the Fermi surface, another singular region of the Green's function occurs near the boundary of the `log-periodic' region of \cite{Faulkner:2009wj}. Here the denominator of the zero temperature Green's function takes the form: ${\mathcal G}(\Omega,  k)^{-1} \sim 1 + e^{-(A+ i B)\lambda(k)} \Omega^{-2 i \lambda(k)}$, where $A$ and $B$ are real constants and $\lambda(k) \to 0$ as $k u_+ \to q \gamma/\sqrt{2}$ (i.e. when $\nu$ pure imaginary goes to $0$ in (\ref{eq:nudef}) below). By explicitly performing the $\Omega$ integral and then bounding the $k$ integral of (\ref{eq:eff4}) in the dangerous small $\Omega$ and small $\lambda(k)$ region, one finds that this region does not lead to singular low temperature behaviour. The effect of temperature can be estimated by replacing $\tanh \frac{\Omega}{2T}$ by an IR cutoff in the frequency integral at $|\Omega| = T$.}
\be
\int_{-\infty}^\infty \frac{|k|dk}{2\pi} \to \int_{-\infty}^{\infty} \frac{|k_F| d k_\perp}{2\pi} \,,
\ee
and set $k \to k_F$ in the remainder of (\ref{eq:eff4}). We will check the self-consistency of this approximation a posteriori.  It is simple to perform the $k_\perp$ integral, leading to
\bea\label{eq:eff5}
F(u,u') & = &  -{2 k_F\over L^6} \, \mbox{Re} \,  \int_{0}^\infty \frac{d \Omega}{\pi} \, \frac{h_1^2 \tanh \frac{\Omega}{2T}}{2 \Omega/v_F + T^{2 \nu} \left( {\mathcal F}_{\nu}(\frac{-\Omega}{T}) - {\overline{{\mathcal F}_{\nu}(\frac{\Omega}{T})}} \right)}\, \times \\
& & \begin{cases}
\psi^\text{bdy}(\Omega, u')^\dagger\psi^\text{bdy}(-\Omega, u')
\psi^\text{in}(\Omega, u)^\dagger\psi^\text{in}(-\Omega, u)  & u>u'\\
\psi^\text{in}(\Omega, u')^\dagger\psi^\text{in}(-\Omega, u')
\psi^\text{bdy}(\Omega, u)^\dagger\psi^\text{bdy}(-\Omega, u)  & u<u'\\
\end{cases}\ .\nonumber
\eea
All wavefunctions in this expression are evaluated at $k = k_F$. In general, as we recall below, there can be multiple Fermi surfaces for a given charge \cite{Liu:2009dm}. Their contributions will sum in the above formula for $F(u,u')$.

The form of the Green's function (\ref{eq:GF}) used in the integral (\ref{eq:eff5}) is only correct for $\Omega \ll \mu$. For our computation to be valid we should therefore make sure that the key contribution to the integral comes from sufficiently low frequencies. More precisely, we will be interested in frequencies in the range $T \ll \Omega \ll  \mu$. This range is the most IR singular, because the $\tanh \frac{\Omega}{2T}$ in the integral can be replaced by $1$, and will be seen to capture the universal pairing physics. We now evaluate the integral for these frequencies.

For $\Omega \gg T$ one can take the low temperature limit (\ref{eq:lowTlimit}) of ${\mathcal F}_{\nu} \left(\frac{\Omega}{T}\right)$. The denominator in (\ref{eq:eff5}) becomes $2 \Omega/v_F + 2 i \, h_2 e^{i \pi \nu}\sin \theta \, \Omega^{2\nu}$. The constant $h_2$ is not dimensionless, but rather $h_2 \sim \mu^{1-2\nu}$. Therefore $T \ll \Omega \ll \mu$ implies, as emphasised in \cite{Faulkner:2009wj}, that the first term ($\sim \Omega$) dominates if $\nu > \half$ while the second ($\sim \Omega^{2\nu}$) dominates when $\nu < \half$. In both cases, the range $T \ll \Omega \ll \mu$ implies that the
fermion wavefunctions should be simply evaluated at $\Omega = 0$ in the extremal $T=0$ black hole background.\footnote{Setting $T=\Omega=0$ and keeping the coordinates of the wavefunctions $u,u'$ finite means that we miss the contribution to the frequency integral from e.g. $1 - u/u_+ \ll T \ll \Omega$. One should worry about the possibility of IR singular temperature dependence arising from this very near horizon region. In an appendix we check that this region does not give additional low temperature divergences to those discussed in the main text.}
Furthermore, given that we are then at $\Omega=0$ and $k=k_F$ (at $T=0$), the wavefunction is precisely the Fermi surface zero mode,
\be
\psi^0 \equiv \psi^\text{bdy}(\Omega=0,k=k_F) = \psi^\text{in}(\Omega=0,k=k_F) \ .
\ee
Let us consider the cases $\nu > \half$ and $\nu < \half$ in turn. 
\begin{itemize}

\item $\nu > \half$:  The leading behaviour of (\ref{eq:eff5}) in an expansion in $T \ll \mu_\text{RG} \lesssim \mu$ is
\be\label{eq:bcslike}
F(u,u') = -\frac{ h_1^2 v_F k_F}{\pi L^6} \left( \log \frac{\mu_\text{RG}}{T} + \g_E + \log \frac{2}{\pi} \right)
\psi^0(u)^\dagger \psi^0(u) \psi^0(u')^\dagger \psi^0(u')  \,,
\ee
where $\g_E$ is Euler's constant.

\item $\nu < \half$:  The leading behaviour of (\ref{eq:eff5}) in an expansion in $T \ll \mu_\text{RG} \lesssim \mu$ is now
\be\label{eq:nonbcslike}
F(u,u') = \frac{ h_1^2 k_F T^{1-2\nu}}{\pi h_2 L^6} \frac{\sin \pi \nu}{\sin \theta} \left(\frac{1}{1 - 2 \nu} \left(\frac{\mu_\text{RG}}{T}\right)^{1-2\nu}+  c_\nu  \right) \psi^0(u)^\dagger \psi^0(u) \psi^0(u')^\dagger \psi^0(u') \,,
\ee
where the constant appearing in the second term is
\be
c_\nu = 2 (2^{2\nu}-1) \Gamma(1 - 2\nu) \zeta(1-2\nu) \,.
\ee
\end{itemize}
In these expressions $\mu_\text{RG}$ is a renormalisation scale which we implement as a hard cutoff on the frequency integral. In any case what will be most important is the temperature dependence and, in the former case (\ref{eq:bcslike}), the coefficient of the logarithm.

The first of the two cases above is particularly interesting. There is a logarithmic divergence associated with low temperatures. The expression in (\ref{eq:bcslike}) is essentially the same as that appearing in BCS theory. It is this first case, $\nu > \half$, in which we will be able to consistently describe the onset of superconductivity. The logarithmic divergence in this case indicates that the range of frequencies and momenta we have integrated over do indeed pick out the dominant contribution to the integrals at low temperatures.

In the second case there is no low temperature divergence. Recall that the origin of superconductivity is the marginal relevancy of the BCS coupling about the Fermi surface \cite{Polchinski:1992ed}. What has happenned in the second case is that the modified (non-Fermi liquid) low energy dispersion relation, $\Omega^{2 \nu} \sim k_\perp$ as opposed to $\Omega \sim k_\perp$, has resulted in the quartic coupling becoming irrelevant and hence not leading to strong IR effects. This is familiar from the bosonic case \cite{hertz}. The absence of a low $T$ divergence means that (\ref{eq:nonbcslike}) is not a controlled approximation to the integral and one should consider the full problem before concluding beyond doubt that there is no superconductivity in this case. We have shown however that there is no instability near the Fermi surface. Consistent with this observation we will find below that $T_c \to 0$ as $\nu = \half$ is approached from above. In some contexts, see e.g. \cite{longrange1, longrange2, longrange3}
for a sampling including a color superconductivity case, long range pairing interactions lead to additional inverse frequency dependence in the analogue of (\ref{eq:eff5}). Such interactions can compensate for the weaker frequency dependence of the fermion propagator, allowing for non-Fermi liquid pairing to occur. In the holographic models under consideration, a nontrivial fermion propagator can result from classical propagation on a curved spacetime, but the leading fermion interactions are mediated by a contact interaction that seems not to allow the non-Fermi liquids to pair.

In both of the above expressions, (\ref{eq:bcslike}) and (\ref{eq:nonbcslike}), the $u$ and $u'$ dependence of $F(u,u')$ has factorised. This allows us to solve explicitly for the unstable mode and the critical temperature $T_c$. The critical temperature is defined by the appearance of a zero mode for the condensate. From the quadratic effective action (\ref{eq:effectiveaction}) this requires
\be\label{eq:eom}
M_F^2 \Delta(u) + \int du' \sqrt{-g(u')}\Delta(u') F(u',u) = 0 \,.
\ee
The factorisation of $F(u',u)$ immediately allows us to conclude that the critical zero mode
\be\label{eq:deltazero}
\Delta^0(u) \propto   \psi^0(u)^\dagger \psi^0(u) \,,
\ee
which seems rather natural. Plugging back into (\ref{eq:eom}) leads to a simple formula for the critical temperature (for $\nu > \half$)
\be\label{eq:tc1}
M_F^2 = \frac{h_1^2 v_F k_F}{\pi L^6} \left(\log \frac{\mu_\text{RG}}{T_c} + \gamma_E + \log \frac{2}{\pi} \right) \int du \sqrt{-g}(\psi^{0\dagger}\psi^0)^2 \,.
\ee
Solving for the critical temperature gives
\be\label{eq:Tc2}
T_c = {\textstyle \frac{2}{\pi}} e^{\gamma_E} \mu_\text{RG} e^{- M_F^2 L^2/N_\text{eff.}(\gamma q)} \,,
\ee
where the effective density of states at the Fermi surface $N_\text{eff.}(\gamma q)$ is a dimensionless function of the fermion charge $q$ in units of $\gamma$, as defined in (\ref{eq:gamma}).
The exponent is the most important part of this expression. In the remainder of the paper we will determine this dependence of $T_c$ on $\gamma q$, which is a free parameter in our theory. For our various approximations to be reliable we need $T_c \ll \mu_\text{RG}$. From (\ref{eq:Tc2}) this is seen to hold when $\frac{N_\text{eff.}(\gamma q)}{M_F^2 L^2} \ll 1$. This last inequality can also be thought of as the condition for validity of perturbation theory in the quartic fermion interaction of our theory (\ref{intera}). That said, the computation of $T_c$ is balancing a classical and one loop mass term at a nonperturbatively (in the coupling) low temperature. Around and below this temperature a marginally relevant coupling is becoming strong and one might worry about the need to resum large logarithms at higher orders in perturbation theory. The one loop computation we have performed is in fact the only fermion loop contribution to the quadratic effective action for $\Delta$ within our theory and so our computation is exact close to the critical temperature. More generally, moving away from the quadratic level, in BCS-Eliashberg theory the kinematics of the Fermi surface leads to the absence of further relevant operators and the one loop computation is exact, see e.g. \cite{Polchinski:1992ed}. It is likely that a similar statement holds for our setup.

As well as the $\g q$ dependence in the exponent, there is the dependence on $\frac{1}{M_F^2 L^2}$. The Fermi mass $M_F$ does not correspond to a dimensionful scale in the dual field theory (the only scales in the otherwise conformal field theory are $T$ and $\mu$). We will discuss in the final section how $\frac{1}{M_F^2 L^2}$ is instead related to a dimensionless four point correlator in the field theory. This correlator therefore controls $T_c$ in our (dual) strongly interacting field theory in the same way the electron-glue coupling controls the critical temperature in a perturbative BCS treatment.

In the prefactor in (\ref{eq:Tc2}) the overall scale is set by $\mu_\text{RG}$. This is a renormalisation rather than physical scale, highlighting the fact that the prefactor is not well defined but is scheme dependent. Presumably $T_c \sim \mu$, as $\mu$ is the only scale in the theory at low temperatures. Renormalisation ambiguities will cancel in  dimensionless quantities such as the ratio of $T_c$ to the zero temperature mass gap. These ambiguities do not
affect the coefficient of the logarithmic divergence at low temperatures, due to frequencies $T \ll \Omega \ll \mu$, but will shift the order one term in (\ref{eq:bcslike}). Our main interest will be the function $N_\text{eff.}(\gamma q)$ in the exponent, which is robust. We have primarily kept the prefactor $\frac{2}{\pi} e^{\gamma_E} \approx 1.13$ in (\ref{eq:Tc2}) in order to emphasize the strong similarity with the standard BCS result.  The only difference between (\ref{eq:tc1}) and the analogous expression appearing in flat space BCS theory is the integral over a spatially dependent zero mode determining the effective density of states at the dual field theory Fermi surface.

As noted above, for large enough $q$ there will be multiple Fermi surfaces.  The condensate (\ref{eq:deltazero}) and critical temperature (\ref{eq:Tc2}) were given for the case of a single Fermi surface, but can be readily generalized to include the contribution from all the surfaces at once.  Each Fermi surface contributes a term of the form (\ref{eq:bcslike}) to $F(u,u')$. Labeling the Fermi surfaces with $\nu>\half$ by $n=1,2,3,\dots$, with corresponding zero modes $\psi^0_{(n)}(u)$, the solution of the integral equation (\ref{eq:eom}) has the form
\be\label{multicon}
\Delta^0(u) = \sum_n \alpha_n \psi^0_{(n)}(u)^\dagger \psi^0_{(n)}(u) \ .
\ee
Plugging this ansatz into the integral equation gives an eigenvalue equation for $T_c$. This can be written as
\be\label{multid}
\det\left[N_\text{eff.}^\text{total} B_{mn} -  C_{mn}\right] =  0 \,,
\ee
with
\bea\label{multidtwo}
B_{mn} &=& \delta_{mn} {\pi L^4\over h_{1}^{(n)2} v_F^{(n)} k_F^{(n)}} \,, \notag\\
C_{mn} &=& \int \sqrt{-g}\psi^0_{(m)}(u)^\dagger\psi^0_{(m)}(u)\psi^0_{(n)}(u)^\dagger\psi^0_{(n)}(u)  \ . \notag
\eea
The critical temperature is then given by
\be\label{eq:tctotal}
T_c = {2\over \pi}e^{\gamma_E}\mu_{RG}e^{-M_F^2L^2/N_\text{eff.}^\text{total}} \,,
\ee
where $N_\text{eff.}^\text{total}$ is the largest eigenvalue of (\ref{multid}).  The corresponding eigenvector is $\alpha_n$, the relative contribution of each zero mode to $\Delta^0(u)$.

We now turn to the computation of the various quantities appearing in (\ref{eq:tc1}). We have been able to find the zero mode
$\psi^0$ analytically when $m=0$, allowing many explicit results.

\section{Solution of the massless Dirac equation}

Consider a massive, charged fermion in the black hole background with the wavefunction
\be\label{defpsi}
\psi = e^{-i\Omega t + i k x}\left(\chi_+ + \chi_-, i(\chi_+ - \chi_-),0,0\right) \ .
\ee
The combinations $\chi_\pm$ are chosen for convenience. As discussed above, the lower two components of the spinor decouple and so can be set to zero. The Dirac equation (\ref{radialdiracmom}) is
\be\label{eq:sepdirac}
D_\mp \chi_\pm = (-{Lm\over u} \pm ik)\sqrt{f}\chi_\mp \,,
\ee
where
\be
D_\pm = f \pa_u - {3f\over 2u} + {f'\over 4} \pm i(\Omega + q A_t) \ .
\ee
The Dirac equation is actually real, and the $i$ appears in (\ref{eq:sepdirac}) only because it was inserted by hand in (\ref{defpsi}).  Generally, this equation must be solved numerically. We will treat the massless case, where we have been able to obtain the wavefunctions analytically for small $\Omega$.  Setting $m=0$, the components of (\ref{eq:sepdirac}) can be decoupled to give the second order equations
\be\label{decoupleddirac}
f \chi_\pm'' + \left(f' - {3f\over u}\right) \chi_\pm' + V_\pm \chi_\pm = 0 \,,
\ee
where
\be
V_\pm = {1\over f}\left[\Omega + qA_t \pm {if'\over 4}\right]^2 + {15 f\over 4 u^2} - {3 f'\over 2 u} + {f''\over 4} \pm {iq\mu\over u_+}-k^2 \ .
\ee

\subsection{Zero modes at zero temperature}\label{s:zeromodes}

Our first objective is to find the fermion wavefunctions at $\Omega = 0$ in the $T=0$ background.
From this point on we restrict to the massless $m=0$ case as we are able to find an analytic solution here.
Define dimensionless quantities and radial coordinate $z$ by
\be\label{eq:rescale}
\tilde{\omega} = \Omega u_+ \ , \quad \tilde{q} = q \mu u_+ \ , \quad \tilde{k} = k u_+ \ , \quad u = u_+(1-z) \ .
\ee
Recall from (\ref{eq:T}) that at zero temperature $u_+ = \sqrt{3} \gamma/\mu$.

Near the horizon, $z \sim 0$, the wavefunctions behave as $z^{-\half \pm \nu_k}$ with
\be\label{eq:nudef}
\nu_k = {1\over 6}\sqrt{6\tilde{k}^2 - \tilde{q}^2} \,.
\ee
We will later be interested in a certain $k = k_F$ and define
\be
\nu \equiv \nu_{k_F} \,.
\ee
This is the $\nu$ we referred to in the previous section.
We will mainly be interested in cases with $\nu > \half$. We require the $z^{-\half + \nu}$ behavior for regularity.

At zero temperature, the function that appears in the metric can be written as
\be
f = 3z^2(z-z_0)(z-\bar{z}_0) \ , \quad z_0 \equiv {1\over 3}(4 + i \sqrt{2}) \ .
\ee
Plugging this into the decoupled equations (\ref{decoupleddirac}) with $\Omega = 0$ gives the equation for the fermion zero modes.  It is found to have the exact solution
\bea\label{zerom}
\lefteqn{\chi_{\pm}^0 = N_\pm (z-1)^{3/2}z^{-\half+\nu_k}(z-z_0)^{-\half-\nu_k}\left(z-\bar{z}_0\over z - z_0\right)^{{1\over 4}(-1 \pm \sqrt{2} \tilde{q}/\bar{z}_0)}} \\
&&  \qquad \times \ _2F_1\left(\half + \nu_k  \pm {\sqrt{2}\over 3}\tilde{q},  \nu_k \pm i{\tilde{q}\over 6}, 1 + 2 \nu_k , {-2i\sqrt{2}z\over 3 \bar{z}_0(z - z_0)}\right) \,, \nonumber
\eea
with $N_\pm$ a normalization.
The second independent solution is obtained by replacing $\nu_k \to - \nu_k$ in (\ref{zerom}),
\be\label{zeroeta}
\eta_{\pm}^0 = \widetilde{N}_\pm \times \left( {\chi_\pm^0\over N_\pm} \ \mbox{with \ } \nu_k \to -\nu_k\right) \,,
\ee
with a new normalization $\widetilde{N}_\pm$. The first solution (\ref{zerom}) has the required regular behavior at the horizon for the $\Omega=0$ solution.  The second will also be required when we consider a small nonzero frequency. Inserting the solution into the first order Dirac equation (\ref{eq:sepdirac}) gives the relative normalizations
\be\label{relnorm}
{N_-\over N_+} = {6 i \nu_k -\tilde{q}\over \tilde{k}\sqrt{6}}\left(\bar{z}_0\over z_0\right)^{\tilde{q}/\sqrt{2} \bar{z}_0} \ , \quad {\widetilde{N}_-\over \widetilde{N}_+} = -{6 i \nu_k +\tilde{q}\over \tilde{k}\sqrt{6}}\left(\bar{z}_0\over z_0\right)^{\tilde{q}/\sqrt{2} \bar{z}_0} \,.
\ee
Thus we have obtained the zero modes $\psi^0$ appearing in the expression (\ref{eq:tc1}) for the critical temperature. While computing $h_1$ and $v_F$ will require moving to small frequencies, we already have enough information to obtain $k_F$.

\subsection{Fermi momentum $k_F$}

\begin{figure}[h]
\begin{centering}
\includegraphics{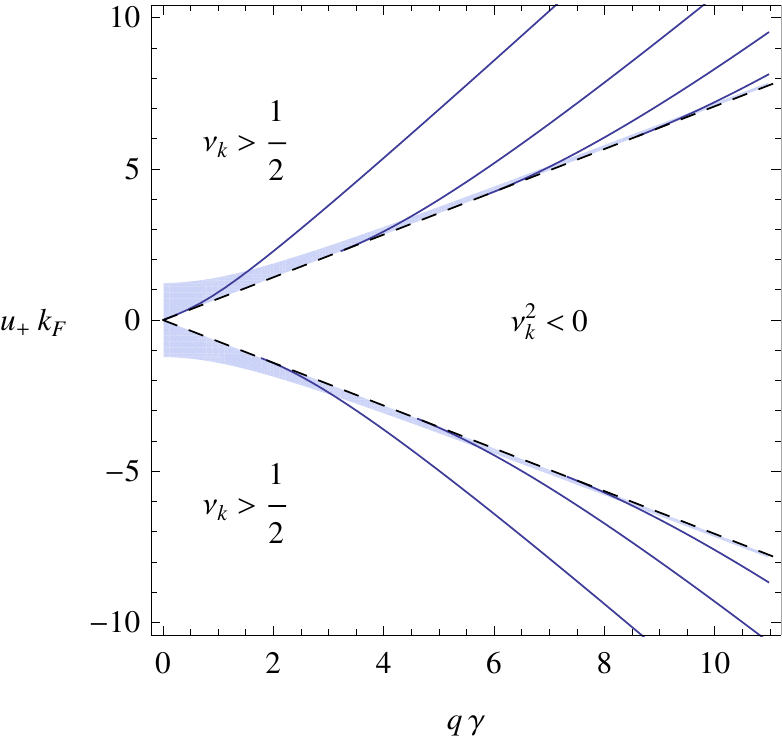}\\
\caption{\small Fermi momentum vs charge of the fermion field.  The dashed line is $\nu_k = 0$ and the shaded region is $0 < \nu_k < \half$. From left to right are the first, second, etc. Fermi surfaces, which disappear when they hit the dashed line.  For the vertical axis, recall that $u_+ = \sqrt{3} \gamma/\mu$. Positive and negative $k_F$ correspond to Fermi surfaces in the Green's functions ${\mathcal{G}}_1$ and ${\mathcal{G}}_2$ respectively.} \label{fig:kf}
\end{centering}
\end{figure}

The Fermi surface is characterized by a zero mode that is regular at the horizon and obeys certain falloff conditions near the boundary of AdS$_4$. As described in Section \ref{sec:retarded}, the asymptotic boundary condition on the fluctuating mode is
\be
\psi^0 = \left(\!
         \begin{array}{c}
           \chi_+^0 +\chi_-^0 \\
           i(\chi_+^0 - \chi_-^0) \\
         \end{array}
       \!\right)
\propto (1-z)^{3/2}
    \left(\!
         \begin{array}{c}
           0 \\
           1 \\
         \end{array}
    \!\right) + \cdots \,.
\ee
The equation for the Fermi momentum $k_F$ is therefore
\be\label{fermizero}
\lim_{z\to 1} (z-1)^{-3/2}(\chi_+^0 + \chi_-^0) = 0 \ .
\ee
Using (\ref{zerom}) and (\ref{relnorm}) in (\ref{fermizero}) gives
\be\label{zeroeq}
{_2F_1\left(1 + \nu - {i\tilde{q}\over 6}, \half + \nu - {\sqrt{2}\tilde{q}\over 3}, 1 + 2\nu, {2\over 3}(1 + i \sqrt{2})\right) \over _2F_1\left( \nu - {i\tilde{q}\over 6}, \half + \nu - {\sqrt{2}\tilde{q}\over 3}, 1 + 2\nu, {2\over 3}(1 + i \sqrt{2})\right)} =  {6\nu + i \tilde{q}\over \tilde{k}_F(2i+\sqrt{2}) } \,.
\ee
The hypergeometric functions can be evaluated numerically to solve for the Fermi surface.  For example, when $\gamma q = 1$, we find $\tilde{k}_F \approx .918528$ in agreement with the numerical solution of the Dirac equation in \cite{Liu:2009dm}.  The solutions of (\ref{zeroeq}) are plotted in figure \ref{fig:kf}. For a given $q$ there can be multiple Fermi surfaces \cite{Liu:2009dm}.  When a distinction is necessary, the largest $|k_F|$ will be called the `first' Fermi surface, the next $|k_F|$ the `second' Fermi surface, and so on.

We can also use the explicit solution above to make an observation about the Green's function at $\Omega=0$.
The boundary Green's function defined in (\ref{wavenormalization}), evaluated at zero frequency, is
\be
\mathcal{G}(\Omega=0, k) = i \lim_{z\to 1}{\chi_+^0 - \chi_-^0\over \chi_+^0 + \chi_-^0} \ .
\ee
We will now show that the imaginary part of this Green's function vanishes when $\nu_k$ is real. For imaginary $\nu_k$
the imaginary part of the Green's function will be positive provided that $i \nu_k$ is taken to be positive (this amounts to a choice of sign of a square root in the above).
For real $\nu_k$, as we are interested in, the wavefunctions can be shown to satisfy
\be
\chi_-^0 = e^{iP} \bar{\chi}_+^0 \,,
\ee
with $P$ the real phase
\be
P = \pi+2\pi\nu_k -  \tan^{-1}\left(6\nu_k\over \tilde{q}\right) - 2 \tan^{-1}\left(\sqrt{2}\over 4\right)(\nu_k + \sqrt{2}\tilde{q}/3) \ .
\ee
This expression assumes for concreteness that $N_+$ is real and $q>0$. It follows that for real $\nu_k$ the spectral density vanishes away from $k=k_F$, i.e. $\mbox{Im} \mathcal{G}(\omega=0,k) = 0$, while the real part has a pole at the Fermi momentum.

By substituting the Fermi momentum into the zero mode wavefunction (\ref{zerom}) we can plot the radial profile of the unstable Cooper pair mode $\Delta^{0}$ of equation (\ref{eq:deltazero}). The result for various charges is shown in figure \ref{fig:condplot} below. The most notable feature of these plots is that for large charge the zero modes are supported away from the horizon while as $\nu \to \half$, at smaller charge, the wavefunctions are supported in the near horizon region. This is consistent with the observation, below and in \cite{Faulkner:2009wj, Faulkner:2010tq}, that for $\nu < \half$ the physics of the (non-Fermi liquid) Fermi surface is captured by the near horizon $AdS_2 \times \R^2$ region.

\begin{figure}[h]
\begin{centering}
\includegraphics[height=5cm]{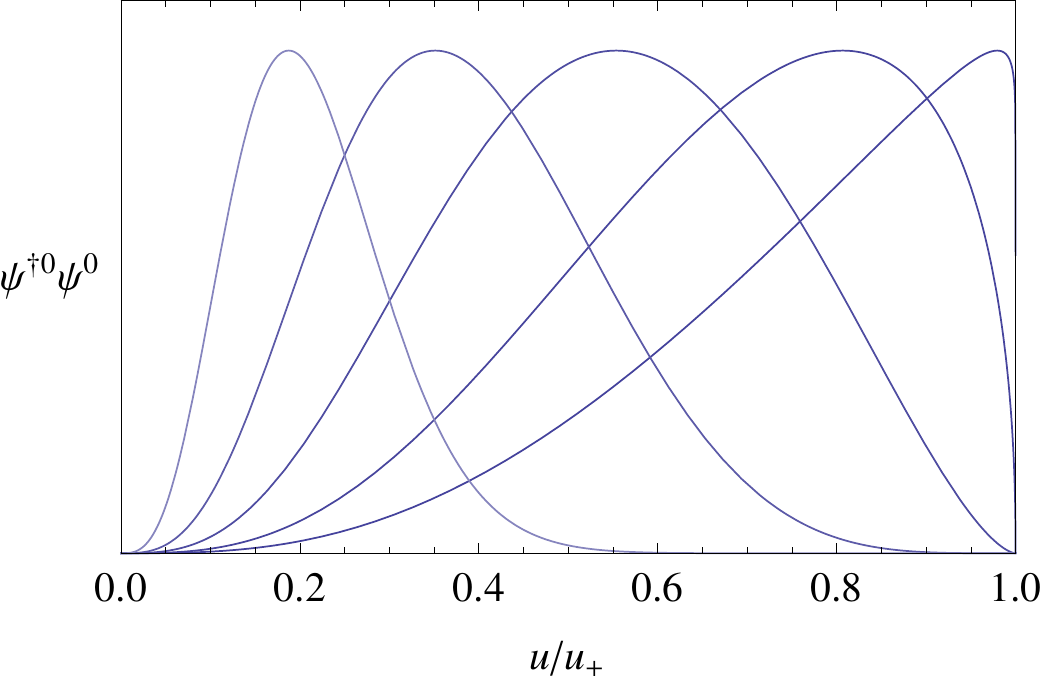}\\
\caption{\small Wavefunctions of the unstable mode $\Delta^{0}$ at different values of the charge for the first Fermi surface. From left to right $\gamma q = \{11, 5, 3, 2, 1.6 \}$. The curves are normalised so that their maxima are equal. At large charge the wavefunctions are supported away from the horizon. As $\nu \to \half$ at smaller charge ($\gamma q \to 1.56$), the wavefunctions are increasingly supported in the near horizon region.} \label{fig:condplot}
\end{centering}
\end{figure}

\subsection{Small frequencies}

Our next goal is to find wavefunctions at $T=0$ for small but nonzero frequencies $\Omega$, using a matching procedure. This was done in \cite{Faulkner:2009wj}, although we now have recourse to the exact zero mode (\ref{zerom}).  We solve the wave equation analytically in the ``near region" defined by $z \ll 1$ and in the ``far region" defined by $z \gg \tilde \w$ (at fixed $\tilde q$).  Matching the two solutions gives a solution on the full spacetime as long as the near and far regions overlap,
\be
\tilde \omega \ll 1 \,.
\ee

\subsubsection{Near region}

In the near horizon region
\be
f = 6 z^2 +O(z^3)\ .
\ee
Similarly expanding all quantities for small $z$ and, anticipating the matching, working to lowest order in $\tilde \w$ in each power of $z$,
the second order wave equation becomes
\be
z^2 \chi_\pm '' +2z \chi_\pm' + {1\over 36}\left( {(\tilde{\omega}+\tilde{q} z)^2\over z^2} \mp {6i\tilde{\omega}\over z } + 9 - 6\tilde{k}^2\right)\chi_\pm = 0 \,,
\ee
with solution
\be
\chi_\pm^\text{near} = A_\pm z^{-\half-\nu_k} e^{-{i\tilde{\omega}\over 6z}} \ _1F_1\left(\half \pm \half + \nu_k + {i\tilde{q}\over 6}, 1 + 2\nu_k, {i \tilde{\omega}\over 3 z}\right) + B_\pm(\nu_k\to-\nu_k) \,.
\ee
On this solution we must also impose the constraint that it satisfies the first order Dirac equation. We will do this in the following subsection as part of matching to the asymptotic `far' solution.

Requiring the solution to have ingoing group velocity at the horizon, i.e. leading behaviour of the form $e^{+i\tilde \w/6z}$, fixes
\be\label{gir}
\mathcal{G}^\text{IR} \equiv {A_+\over B_+} =     { \Gamma(-2\nu_k)\Gamma(\nu_k -  {i\tilde{q}\over 6})\over 9^{\nu_k} \Gamma(2\nu_k)\Gamma(-\nu_k - {i \tilde{q}\over 6})}(- i\tilde{\omega})^{2\nu_k} \, ,
\ee
and similarly for $A_-/B_-$.
The ratio $\mathcal{G}^\text{IR} = A_+/B_+$ is the two-point function in the IR CFT living on the boundary of AdS$_2$. In the matching region $z \gg \tilde{\omega}$,
\be\label{nearmatching}
\chi_\pm^\text{near} = A_\pm z^{-\half - \nu_k} + B_\pm z^{-\half + \nu_k} \,.
\ee

\subsubsection{Far region}

In the asymptotic `far' region $z \gg \tilde \omega$, we solve the wave equation perturbatively in $\tilde \omega$, expanding
\be\label{eq:ff}
\chi_\pm^\text{far} = \chi_\pm^{(0)} + \tilde{\omega} \chi_\pm^{(1)} + \cdots \,,
\ee
$\chi_\pm^{(0)}$ is a combination of the zero modes found in Section (\ref{s:zeromodes}),
\be
\chi_\pm^{(0)} = \chi_\pm^0 + \eta_\pm^0 \,,
\ee
where $\chi_\pm^0$ and $\eta_\pm^0$ were defined in (\ref{zerom}), (\ref{zeroeta}), and have normalizations $N_\pm$ and $\widetilde{N}_\pm$ respectively. Expanding in the matching region $z \ll 1$,
\be\label{farmatching}
\chi_\pm^\text{far} =\widetilde{N}_\pm S^\pm_\nu z^{-\half - \nu} + N_\pm S^\pm_{-\nu} z^{-\half + \nu}  + \ocal(\tilde{\omega}) \,,
\ee
with
\be\label{eq:ss}
S^\pm_\nu = (-1)^{3/2}(-z_0)^{-\half + \nu_k}\left(\bar{z}_0\over z_0\right)^{-{1\over 4} \pm {\sqrt{2}\tilde{q}\over 4 \bar{z}_0}} \,.
\ee
Comparing the near solution (\ref{nearmatching}) and far solution (\ref{farmatching}) in the matching region $\tilde \omega \ll z \ll 1$ determines the relative contribution of $\chi^0_\pm$ and $\eta^0_\pm$ in the far region,
\be\label{matchsol}
{\widetilde{N}_+ \over N_+} = (-z_0)^{-2\nu_k} \mathcal{G}^\text{IR} \ .
\ee
Recall that $N_-$ and $\widetilde N_-$ were given in terms of $N_+$ and $\widetilde N_+$ in (\ref{relnorm}) above. Note that $\mathcal{G}^\text{IR}$ obtained in (\ref{gir}) is frequency dependent. We will mostly be interested in $\nu > \half$ and so this frequency dependence is subdominant compared to the  order $\tilde \w$ term in (\ref{eq:ff}) and (\ref{farmatching}), which we turn to now.

The first order correction $\chi_\pm^{(1)}$ satisfies an inhomogeneous wave equation with $\chi_\pm^{(0)}$ as the source.  We will need only the leading asymptotic behavior near the boundary $z \to 1$. The asymptotic behaviour can be found elegantly by integrating the Dirac equation (\ref{eq:sepdirac}) as in Appendix C of \cite{Faulkner:2009wj}. To leading order at the boundary $z \to 1$ one finds
\be\label{v1}
\chi_+^{(1)} + \chi_-^{(1)} = -(1-z)^{3} {2 i u_+^2\over L^3}{\int_0^{u_+} du \sqrt{g g^{tt}} \left(|\chi_+^0|^2 + |\chi_-^0|^2\right) \over  \chi_+^{0*} - \chi_-^{0*}}  + \cdots \,.
\ee
In obtaining this expression there is a contribution at the horizon which vanishes in the cases of interest to us ($\nu > \half$), see \cite{Faulkner:2009wj}.

\subsubsection{Boundary Green's function}

\begin{figure}[h]
\begin{centering}
\includegraphics{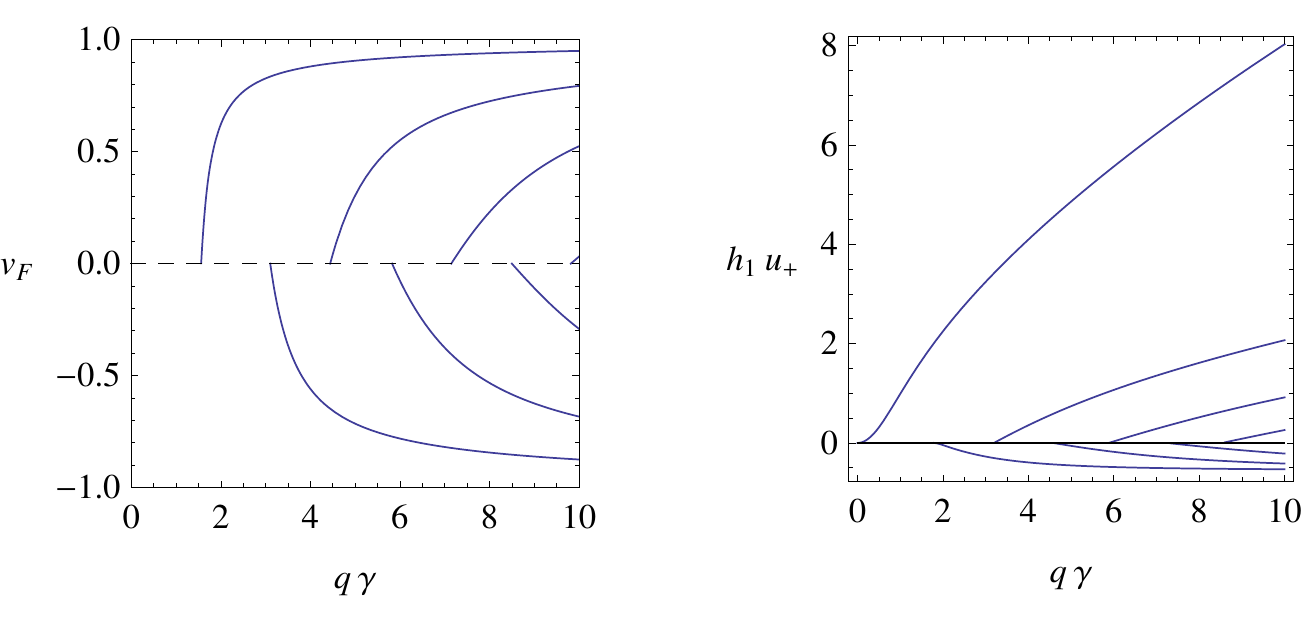}\\
\caption{\small  Left: Fermi velocity vs charge, equation (\ref{vequation}). $v_F$ vanishes at the dashed line when $\nu=\half$. Right: $h_1$ vs charge, equation (\ref{hequation}). $h_1$ vanishes at the solid line when $\nu=0$. The multiple lines in each plot are for the various Fermi surfaces, in ascending order with the first Fermi surface on the left. Note that $v_F$ and $h_1$ have the same sign as $k_F$. As above, positive and negative $k_F$ correspond to Fermi surfaces in the Green's functions ${\mathcal{G}}_1$ and ${\mathcal{G}}_2$ respectively. \label{fig:multiplot}}
\end{centering}
\end{figure}

Finally we can obtain the Green's function of the dual field theory living on the boundary of AdS$_4$, as defined in equation (\ref{wavenormalization}).
This is given by
\be
\mathcal{G} =  i \lim_{z\to 1} {\chi_+ - \chi_-\over \chi_+ + \chi_-} \ .
\ee
Explicitly in a small $\tilde \w$ expansion:
\be
\mathcal{G} = i \lim_{z\to 1} {\chi_+^0 - \chi_-^0 + \eta_+^0 - \eta_-^0 + \mathcal{O}(\tilde{\omega})
\over \chi_+^0 + \chi_-^0 + \eta_+^0 + \eta_-^0 + \tilde{\omega}(\chi_+^{(1)} + \chi_-^{(1)})+ \mathcal{O}(\tilde{\omega}^2)} \,,
\ee
where the wavefunctions $\chi_\pm^0$, $\eta_\pm^0$ defined in (\ref{zerom}, \ref{zeroeta}) are normalized according to (\ref{relnorm}, \ref{matchsol}), and the asymptotic behavior of the last term in the denominator was given in (\ref{v1}). Note that due to the matching condition (\ref{matchsol}), \be
\eta_\pm^0 \propto \mathcal{G}^\text{IR} \propto \tilde{\omega}^{2\nu_k} \ ,
\ee
while all other wavefunctions are independent of $\tilde\omega$.
Near the Fermi surface $k=k_F+k_\perp$, the Green's function becomes
\be
\mathcal{G} \approx {h_1\over k_\perp - \Omega/v_F + h_2 e^{i\theta-i\pi\nu}\Omega^{2\nu}} \,,
\ee
which is the advertised zero temperature limit of (\ref{eq:GF}), using (\ref{eq:lowTlimit}). In this above formula
\bea\label{hequation}
h_1 &=& {i\over u_+}\lim_{z\to 1}{\chi_+^0 - \chi_-^0\over \pa_{\tilde{k}}(\chi_+^0 + \chi_-^0)} \,, \\
v_F &=& {L^3\over u_+^3 h_1}\left(\int_0^{u_+} du \sqrt{g g^{tt}} \psi^{0\dagger}\psi^0\right)^{-1} \lim_{z\to 1}(1-z)^{-3}|\chi_+^0 - \chi_-^0|^2 \ ,\label{vequation}
\eea
with all wavefunctions evaluated at $k = k_F$. $\psi^0$ is the zero mode spinor with components $\chi_\pm^0$ as in (\ref{defpsi}). $h_2,\theta$ are easily obtained from the expression above but will not be needed. Note that all of these wavefunctions are hypergeometric functions that have been given analytically above. From these expressions one finds that $h_1$ and $v_F$ are real. These quantities are plotted as a function of charge in figure \ref{fig:multiplot}. Similar plots for the first and second Fermi surfaces are found in \cite{Faulkner:2009wj}.

The decrease of the Fermi velocity with charge and the fact that the velocity tends to the speed of light at large charges in figure \ref{fig:multiplot} suggest a geometrical interpretation. As the charge is lowered, the zero mode wavefunction is supported increasingly close to the black hole horizon, c.f. figure \ref{fig:condplot}. The gravitational redshift then reduces the local speed of light relative to the boundary value. This observation is made in passing in \cite{Faulkner:2009wj}, while similar phenomena have been discussed previously in different contexts, e.g. \cite{Mateos:2007yp}.

\subsection{The critical temperature as a function of charge}

We can now obtain the critical temperature from (\ref{eq:Tc2}), which we rewrite here as
\be\label{eq:Tc3}
T_c \propto \mu e^{- M_F^2 L^2/N_\text{eff.}} \,,
\ee
with
\be\label{eq:n}
N_\text{eff.} = \frac{ h_1^2 v_F k_F}{\pi L^4} \int_0^{u_+} du \sqrt{-g} (\psi^{0\dagger} \psi^0)^2 \,.
\ee
Using our zero mode (\ref{zerom}) as well as the above results for $\{h_1, v_F, k_F\}$ we plot $N_\text{eff.}$ below in figure \ref{fig:Neff}. The zero mode should be normalised overall according to (\ref{wavenormalization}). As well as the $N_\text{eff.}$ for each Fermi surface individually there is also the total effective density of states $N_\text{eff.}^\text{total}$ and corresponding critical temperature $T_c$ of (\ref{eq:tctotal}). Thus $N_\text{eff.}^\text{total}$ determines the actual critical temperature of the system and is also plotted.

\begin{figure}[h]
\begin{centering}
\includegraphics[height=6.5cm]{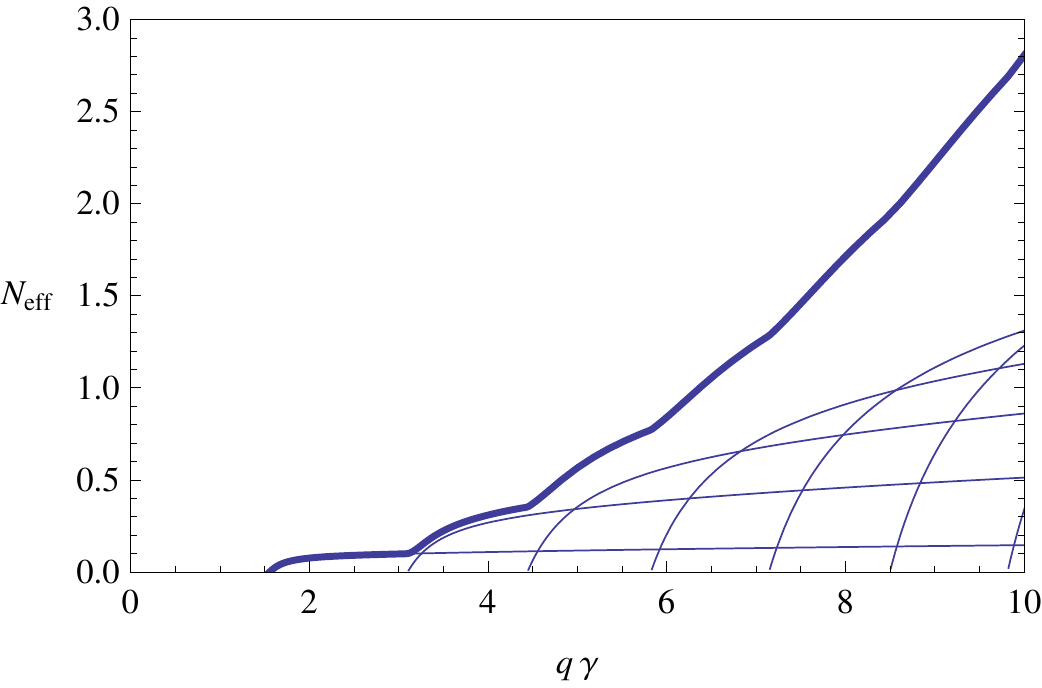}\\
\caption{\small  Lighter lines show the effective density of states at the Fermi surface $N_\text{eff.}$ vs charge, for each Fermi surface with the first Fermi surface on the left. Recall from (\ref{eq:Tc3}) that $T_c \sim \mu e^{- M_F^2 L^2/N_\text{eff.}}$. The dark line is $N_\text{eff.}^\text{total}$, defined in (\ref{multid}), which accounts for the presence of multiple Fermi surfaces. \label{fig:Neff}}
\end{centering}
\end{figure}

When the density of states $N_\text{eff.}$ goes to zero, then clearly from (\ref{eq:Tc3}) the critical temperature $T_c$ also vanishes. This is intuitively reasonable given that it is the states at the Fermi surface that pair and condense. Similarly, the larger $N_\text{eff.}$, at fixed $M_F^2 L^2$, the larger the critical temperature. Therefore in figure \ref{fig:Neff} we see that the critical temperature goes to zero at $\gamma q \approx 1.56$, corresponding to $\nu = \half$. Also interesting in this plot is that the fermions that pair at the highest temperature can be on different Fermi surfaces at different charges, leading to kinks in $T_c$ as a function of $\gamma q$.

Recall that for our computation to be consistent we need $T_c \ll \mu$. We see in figure \ref{fig:Neff} that for fixed $M_F^2 L^2$ this requirement will break down at very large charge. It would be interesting to relax this condition by working with general charged black holes rather than the near extremal solutions we have considered. In the general case it will likely not be possible to solve the equations analytically.

\section{Discussion and dual field theory}

Our main result is figure \ref{fig:Neff} which shows the critical temperature for the onset of superconductivity as a function of the charge of a massless fermion field in a near extremal
charged black hole background. There are two immediately interesting aspects of this plot. Firstly, that at sufficiently large charge there can be crossovers at which different Fermi surfaces give the highest critical temperatures. Secondly, that the critical temperature goes to zero at the charge such that the critical exponent $\nu = \half$, with apparently no superconducting phase transition from this mechanism for $\nu < \half$.

The regime $\nu \leq \half$ contains the non-Fermi liquids described in \cite{Faulkner:2009wj}, including the marginal case. The lack of pairing in these cases is potentially disappointing from the point of view of dually describing the emergence of nonconventional superconductivity from a non-Fermi liquid, and of thereby modeling interesting experimental systems. It is a posteriori not surprising however: the vanishing $T_c$ is directly related to a vanishing density of states at the Fermi surface which is in turn related to the vanishing of the residue of the Fermi surface pole of the Green's function, which is itself the source of the non-Fermi liquid behaviour. We note below that one could explore alternative pairing mechanisms to attempt to evade this conclusion. Alternatively the onset of superconductivity may not be related to the fermions (at least not explicitly), along the lines of  \cite{Hartnoll:2008vx, Hartnoll:2008kx}.

Before discussing various loose ends in our above computations and indicating future avenues for research, we should characterise the onset of superconductivity from the dual field theory perspective. In the bulk we have essentially repeated the BCS computation in a curved spacetime. The dual field theory is however strongly coupled. The exponent in our formula for the critical temperature (\ref{eq:Tc2}) depends on two parameters, the effective density of states at the Fermi surface, $N_\text{eff.}(\g q)$, and the bulk coupling $\frac{1}{M_F^2 L^2}$. The effective density of states has the same interpretation in the field theory. The bulk coupling however determines the leading contribution to a four-point correlator of the fermionic field. Let $\Psi$ be the (two component) fermionic operator of the dual field theory that is sourced by the boundary value of the (two) non-normalisable components of the bulk fermion $\psi$, as discussed around equation (\ref{wavenormalization}) above. Then
\be
\langle \Psi \Psi \overline \Psi \overline \Psi \rangle \sim \frac{1}{M_F^2 L^2} \ ,
\ee
where a particular spin index structure and coordinate dependence follow from the interaction (\ref{intera}) and could be computed from Witten diagrams. To summarise: the critical temperature is controlled by the magnitude of a specific fermion four point function in the zero temperature, zero chemical potential strongly coupled quantum critical theory. This fact parallels the observation in \cite{Denef:2009tp} that for the bosonic holographic superconductors, the critical temperature is determined by the two point function of the bosonic field. In both cases the answer is phrased in terms of quantities that are natural at strong coupling.

In computing the effective action for the Cooper pairs $\Delta$ we have considered the effect of virtual fermions, but not of photon or graviton loops. These loops will dress the fermion propagator and also renormalise the $\Delta \psi^2$ interaction of our decoupled Lagrangian. From the Einstein-Maxwell action (\ref{eq:einsteinmaxwell}) we see that photon and graviton loops will come with powers of  $g^2$ and $\frac{\k^2}{L^2}$. Often in string theory realisations $g^2 \sim \frac{\k^2}{L^2} \sim N^{- \#} \ll 1$, where $\#$ is a positive number and $N$ is a measure of the (large) number of degrees of freedom per site in the dual field theory. These processes are therefore strongly suppressed compared to the term we have computed. There may be interesting physics associated with the resummation of these terms which deserves further study. While the long range repulsive photon should be screened, the (electric and magnetic) photon `vertex corrections' may lead to important effects if $N$ is sufficiently small. To be completely safe with what we have done here we could take the strict large $N$ limit to set these terms to zero. If we do this we should keep $\frac{1}{M_F^2 L^2}$ finite but small in the large $N$ limit. While this introduces a hierarchy between the matter and gravitational sector, such hierarchies are famously known to arise in actually existing quantum gravity systems.

Various future directions present themselves. Perhaps most interestingly, we have not discussed a specific pairing mechanism in the black hole background that would generate a quartic fermion interaction similar to the one we introduced by hand in equation (\ref{intera}).
While simply introducing a quartic interaction is consistent with a bulk effective field theory approach to holography, it may be that various natural candidates for the glue -- say scalars, overscreened photons, nonabelian fields or gravitons\footnote{The possibility of gravitational pairing was mentioned in \cite{Faulkner:2009am}.} -- lead to distinctive pairing physics with interesting dual field theory interpretations. It would be interesting in this context to make contact with the physics of colour superconductivity \cite{Alford:2007xm, Rajagopal:2000wf}. Presumably one can obtain a condensate with p and d wave pairing. Perhaps different interactions can persuade the reluctant modes with $\nu < \half$ to condense. To explore pairing of the $\nu < \half$ modes, the formalism recently developed by \cite{Faulkner:2010tq} may be useful. We noted in our discussion in the main text above that in other contexts, e.g. \cite{longrange1, longrange2, longrange3}, non-Fermi liquids naturally come with long range interactions (as opposed to our contact interaction) which can induce pairing instabilities in non-Fermi liquids.

We have also not discussed the physics of the Cooper pair condensate below the critical temperature. Natural quantities to compute would be the effective bulk Landau-Ginzburg action just below $T_c$ and the zero temperature energy gap. There are two complications in doing this relative to the usual flat space BCS gap computation. The first is that when the mass of the fermion is of order the curvature of the background spacetime, as is often the case in applied holography, then the effective action for the condensate will be nonlocal in the radial direction. The second difficulty is that often the most interesting zero temperature physics will include the backreaction of the condensate on the geometry. Selfconsistently solving simultaneously for the background and the condensate would involve computing functional determinants in an unspecified background. This latter complication could be avoided by working in a probe limit for the condensate. If the condensate is stabilised by higher order terms in the effective action for $\Delta$ (as opposed to via its interaction with the Maxwell field, as in \cite{Hartnoll:2008vx}) then working in the limit $M^2 L^2 \gg \{ g^2, \frac{\k^2}{L^2} \}$ should be sufficient. In such a probe limit it may be possible to solve the nonlocal gap equation.

It should be possible to adapt our computation to the case of rotating black holes and the Fermi surface discovered in \cite{Hartman:2009qu}. Furthermore, that paper used an (astrophysically applicable) WKB approach to reduce the computation of Green's functions to finding geodesics. In our context this could appear as e.g. a large mass and charge limit of the bulk fermionic fields ($\nu$ can be kept fixed in this limit). It would be interesting to see to what extent such a limit simplifies
our one loop calculation and perhaps gives a handle on the question of backreaction mentioned above, by giving a local effective action.

\section*{Acknowledgements}

During this work we have benefitted from insightful conversations with Dionysios Anninos, Nabil Iqbal, John McGreevy, Max Metlitsky, Subir Sachdev and Andy Strominger. Our research is partially supported by DOE grant DE-FG02-91ER40654 and (S.A.H.) the FQXi foundation.

\appendix

\section{Green's function conventions}\label{app:green}

In this appendix we define our conventions for various bulk Green's functions of the Dirac fermion $\psi$. The Euclidean, retarded and advanced Green's functions are defined by
\bea\label{greencon}
G_E(x,x') &=& -\langle T_E\psi(x)\bar\psi(x')\rangle \,, \\
G_R(x,x') &=& -i\theta(t-t')\langle \{\psi(x), \bar\psi(x')\}\rangle\nonumber \,, \\
G_A(x,x') &=& i\theta(t'-t)\langle \{\psi(x), \bar\psi(x')\}\rangle\nonumber \,,
\eea
where $T$ is real time ordering and $T_E$ orders with respect to Euclidean time $\tau = it$, or more generally, $-\mbox{Im}\, t$. For real frequency $\Omega$, the Fourier transformed advanced and retarded Green's function are easily seen to be related by
\be\label{appgid}
G_A(\Omega, \vec{k}, u, u') = \gammat G_R(\Omega,\vec{k},u',u)^\dagger \gammat \,,
\ee
where the transpose in $G_R^\dagger$ acts on spin indices. We have used $\Gamma^{\underline{t}}=\Gamma^{\underline{t}T}=-\Gamma^{\underline{t}*}=-(\Gamma^{\underline{t}})^{-1}$.  The relative sign conventions in (\ref{greencon}) are chosen so that all of the Green's functions satisfy the same equation:
\be
(\Gamma^\mu D_\mu - m)G = i \delta^{(4)}/\sqrt{-g} \ .
\ee
This follows from $(\gammat)^2 = -1$ and the equal time commutator
\be
\{\psi_a(x), \psi_b^\dagger(x')\} = \delta^{(3)}(x,x')/\sqrt{h} \,,
\ee
with $h$ the metric on a spatial slice.

In the black hole background, using (\ref{gammas}) and (\ref{eq:explicit}), acting on one of the two-component projections we have
\be
-i\sigma^1 G_R(\omega, \vec{k}, u',u)^\dagger i\sigma^1 = G_R(\omega, \vec{k}, u, u')^* \,.
\ee
Therefore from (\ref{appgid})
\be\label{lksa}
G_A(\omega,\vec{k},u,u') = -G_R(\omega, \vec{k}, u, u')^* \,.
\ee
In order to confirm that the convention (\ref{greencon}) corresponds to the $G_{A,R}$ used in the derivation of the effective action, note that regularity at the Euclidean horizon in (\ref{finitethorizon}) uniquely fixes the Green's functions that should be used in the Lorentzian formula.  For $\mbox{Im} z>0$, we must pick the ingoing solution, and for $\mbox{Im} z < 0$, we must pick the outgoing solution.  Therefore the Green's function $G^A$ in the derivation of $F(u,u')$ around (\ref{continueg}) is unambiguously defined to be the Green's function in Lorentzian signature with outgoing boundary conditions.  Therefore $G_A$ is obtained by replacing
\be
\psi^\text{in} \to \psi^\text{out} \,,
\ee
in the retarded Green's function (\ref{eq:explicit}).  From our solution of the Dirac equation, we see that
\be
\psi^\text{out} = \psi^\text{in*} \ .
\ee
Also from the Dirac equation $\psi^\text{bdy}$ is real, since it has a real boundary condition.  Therefore we find
\be
G_A(\omega,k,u,u') = -G_R(\omega,k,u,u')^* \,,
\ee
in agreement with the conventions (\ref{greencon}).

\section{Contributions from the near horizon region}

In the text we have shown that when $\nu > \half$ there is a logarithmic low temperature divergence in
the effective mass for the Cooper pairs, as described by the quantity $F(u, u')$ in
equations (\ref{eq:effectiveaction}) and (\ref{eq:eff4}). In evaluating (\ref{eq:eff4}) we substituted the zero mode at the Fermi surface $\psi^0(u)$ for the wavefunctions $\psi^\text{in/bdy}(\Omega, u)$.
In regions where an expansion in powers of $\Omega$ is possible, this substitution
picks out the most IR singular terms, as higher terms come with additional positive powers of $\Omega$.
However, there are three dimensionful quantities of interest: $\{\Omega,T,u\}$, and the vanishing limits of these quantities do not commute. In particular, putting $T=\Omega=0$, as we have done in the wavefunction,
cuts out the range of integration where $z  \lesssim \Omega u_+$ and also the coordinate range $z  \lesssim T u_+$ (recall that $u = u_+(1-z)$). In this appendix we check that there are no additional logarithmic (or worse) temperature divergences from this near horizon region. To do this we must generalise the discussion in the text to (small) finite temperatures.

In order to study this region, we need the wavefunctions for small $\tilde\w = \Omega u_+,\, \t = 4 \pi T u_+$ and $z$. The second order radial wave equation in this regime for a spinor of the form (\ref{defpsi}) is
\be
z\left(6z+\tau\right)\chi_\pm'' + (12z + \tau)\chi_\pm' + \left(3 - \tilde{k}^2 \pm i \tilde{q} - {[12z + \tau \pm 4 i (\tilde{\omega} + z \tilde{q})]^2\over 16z(6z + \tau)}\right)\chi_\pm  = 0 \ .
\ee
Here $\tilde q$, $\tilde k$ and $\nu$ are as defined in (\ref{eq:rescale}) and (\ref{eq:nudef}).
The ingoing and outgoing solutions to this equation are
\bea
\tilde{\chi}_\pm^\text{in} &=& z^{\mp {1\over 4} - {i\tilde{\omega}\over \tau}}(6z+\tau)^{\mp {1\over 4} - {i\tilde{q}\over 6} + {i\tilde{\omega}\over\tau}}\times\\
 & &_2F_1\left(\half  \mp \half  + \nu_k -  {i\tilde{q}\over 6},\half  \mp \half - \nu_k -  {i\tilde{q}\over 6}, 1 \mp \half - {2i\tilde{\omega}\over\tau}, -{6z\over\tau}\right) \,, \notag\\
\tilde{\chi}_\pm^\text{out} &=& \tilde{\chi}_{\mp}^{\text{in}*} \,.
\eea
The normalized `in' and `boundary' wavefunctions defined in (\ref{wavenormalization}) are linear combinations of these solutions in the near horizon region
\bea\label{normalinb}
\chi_\pm^\text{in} &=& D_\pm \tilde{\chi}_\pm^\text{in} \,, \\
\chi_\pm^\text{bdy} &=& E_\pm \tilde{\chi}_\pm^\text{in} + F_\pm \tilde{\chi}_\pm^\text{out}  \label{normalbdyb}\,,
\eea
where $D,E,F$ are $\omega,T$-dependent normalizations to be determined.

For the ingoing mode, expanding in the matching region ($z \gg \w, T$),
\be
\tilde{\chi}_\pm^\text{in} \propto z^{-\half + \nu} + G^{IR}_\pm z^{-\half-\nu} \,,
\ee
with the finite temperature IR Green's function
\be
G^{IR}_\pm = \left(\tau\over 6\right)^{2\nu_k} {\Gamma(-2\nu_k)\Gamma(\half \mp \half + \nu_k - i {\tilde{q}\over 6})\Gamma(\half + \nu_k + i{\tilde{q}\over 6} - i{2\tilde{\omega}\over \tau}) \over
\Gamma(2\nu_k)\Gamma(\half \mp \half -\nu_k - i {\tilde{q}\over 6})\Gamma(\half - \nu_k + i{\tilde{q}\over 6} - i{2\tilde{\omega}\over \tau})} \ .
\ee
Comparing coefficients to the far region wavefunction (\ref{farmatching}) we find the normalization
\bea
D_{\pm} &=& N_\pm S_{-\nu}^{\pm} {\Gamma(\half \mp \half - {i\tilde{q}\over 6} + \nu_k)\Gamma(\half + \nu_k  + {i\tilde{q}\over 6} - {2 i \tilde{\omega}\over \tau})\over\Gamma(2\nu_k)\Gamma(1 \mp \half - {2 i \tilde{\omega}\over \tau}) 6^{\mp \frac{1}{4} + i\frac{\tilde \w}{\tau}- i \frac{\tilde q}{6}}}\left(\tau\over 6\right)^{-\half \pm \half + {i \tilde{q}\over 6} + \nu_k} \,,
\eea
where $S_\nu^{\pm}$ was defined in (\ref{eq:ss}), $N_+$ is chosen to normalize the zero mode at the boundary such that
\be
\lim_{z\to 1}(1-z)^{-3/2} \chi_+^0 = -{i u_+^{3/2}\over 2} \,,
\ee
and $N_-$ is determined by (\ref{relnorm}). The condition above is the first equality of (\ref{wavenormalization}). The constants $N_\pm$ characterise the $\w=T=0$ solution and so do not depend on $\tilde \omega$ and $\tau$. This fixes the normalization of the `in' mode.

For the `boundary' mode, in the far region we can expand
\be
\psi^\text{bdy} = \psi^\text{in} + O(\tilde{\omega}, k_\perp) \ .
\ee
Therefore the normalizations of $\chi_\pm^\text{bdy}$, to leading order near the Fermi surface, are schematically of the form
\bea
E_\pm &\sim& D_\pm[1 + \tilde{\omega} + k_\perp + (G_\pm^{IR})^{-1}( \tilde\omega +  k_\perp)] \,, \\
F_\pm &\sim& D_\pm[1 + \tilde{\omega} + k_\perp + (G_\pm^{IR})^{-1}( \tilde\omega +  k_\perp)] \ ,\notag
\eea
where all the relative coefficients are independent of $\tilde{\omega}$ and have been suppressed.
The inverse factors of $G_\pm^{IR}$ are present so that the $\tilde \w$ dependent term behaves like $\tilde \omega z^{-\half \pm \nu}$ in the matching region.
Expanded for $\tilde{\omega} \gg \tau$, the regime of interest shortly, the various normalization factors in (\ref{normalinb}) and (\ref{normalbdyb}) have frequency dependence
\be\label{highwnorm}
D_\pm \sim \tilde{\omega}^{-\half \pm \half + {i \tilde{q}\over 6} + \nu_k}  \ , \quad
E_\pm \sim F_\pm \sim D_\pm[1 + \tilde{\omega} + k_\perp + \tilde{\omega}^{-2\nu_k}(\tilde{\omega} + k_\perp)] \ .
\ee

\subsection{Checking for near horizon, low temperature divergences}

Low temperature IR divergences arise in the $\Omega$ integral of (\ref{eq:eff4}) in the range $T \ll \Omega$.
This can be checked explicitly, largely following from the fact that for $\Omega \ll T$ the $\tanh \frac{\Omega}{2T}$ term gives an extra
power of $\Omega$ in the numerator. In any case a divergence for $\Omega \lesssim T$ would indicate an IR divergence that
is not cured by finite temperature, which is not the physics we are after.  In the regime $T \ll \Omega$ we can replace
$\tanh \frac{\Omega}{2T} \to 1$ and $T^{2 \nu} {\mathcal{F}}_\nu(\frac{\Omega}{T}) \to \Omega^{2\nu}$ in (\ref{eq:eff4}).
We will check for IR divergences in the seemingly more dangerous case of $\nu > \half$. Similar computations go through for $\nu < \half$.
For the purposes of isolating possible low temperature divergences, the integral in (\ref{eq:eff4}) may therefore be written (with $z<z'$ without loss of generality).
\be\label{eq:appendF}
F(z,z') \sim \text{Re} \int_\tau \frac{d \tilde \omega}{\tilde \omega} \psi^\text{bdy}(\tilde \omega, z')^\dagger\psi^\text{bdy}(-\tilde \omega, z')
\psi^\text{in}(\tilde \omega, z)^\dagger\psi^\text{in}(-\tilde \omega, z) \,.
\ee
We now need to substitute the low temperature, low frequency, near horizon solutions (\ref{normalinb}), normalized according to (\ref{highwnorm}), into this integral and study the temperature dependence.

The dangerous regime is $z, z' \lesssim \tau \ll \tilde \w$ because if  $\tau \lesssim z, z' \ll \tilde \w$ then the $\tilde \w$ integral is cut off by $z$ or $z'$ rather than $\tau$ and there cannot be a temperature dependent IR divergence\footnote{One should also worry about $z,z' \to 0$ divergences in $F(z,z')$ that could become temperature dependent divergences upon performing the integral in the equation of motion (\ref{eq:eom}). The most dangerous region is $T \ll z' < z \ll \tilde \omega$. Using an argument similar to (\ref{maxbound}) below, we have checked that the small $z,z'$ behaviour does not lead to new low temperature instabilities.}. In the regime $z, z' \lesssim \tau \ll \tilde \w$ one can expand (\ref{normalinb}) to find e.g.
\be
\chi^\text{in}_\pm  \sim  \tilde \w^{-\half \pm \half + \nu + i \frac{q}{6}}
z^{\mp \frac{1}{4} - i \frac{\tilde w}{\tau}} (6 z + \tau)^{\mp \frac{1}{4}-i \frac{q}{6} + i \frac{\tilde \w}{\t}} \,,
\ee
and similarly for $\chi^\text{bdy}_\pm$.
Substituting these expressions into (\ref{eq:appendF}) leads to various terms. The fact that $0 \leq z, z' \lesssim \tau$ means that powers of $z,z'$ introduce factors of the temperature. The only dangerous terms can be seen to take the form
\be\label{eq:cosintegral}
F(z,z') \sim \sqrt{\frac{z}{z'}} \int_\tau \frac{d\tilde \omega}{\tilde \w} \cos \left(\frac{\tilde \w}{2 \t} \log \frac{z}{z'} \frac{6 z' + \t}{6 z + \t} \right) \,.
\ee
When $z=z'$ this integral scales like $\log \t$, while for fixed $z \neq z'$ the zero temperature limit is regular due to strong oscillations of the integrand.
Thus we find that we do indeed have an additional logarithmic low temperature divergence due to the near horizon region. However, because this divergence is restricted to the small region $0 \leq z, z' \lesssim \tau$, we can now immediately see that it does not affect our results in the main text. In particular, it does not lead to a new instability localised in the near horizon region: The equation of motion for the zero mode (\ref{eq:eom})
requires
\be\label{maxbound}
\max_{z} \left| \Delta(z) \right| \sim \frac{1}{M^2_F L^2} \max_z \left| \int^\tau dz' F(z,z') \Delta(z') \right| \lesssim \frac{\t \log \t}{M_F^2 L^2} \max_{z} \left| \Delta(z) \right| \,.
\ee
In the perturbative limit $M^2_F L^2 \gg 1$ that we are considering, this equation does not have any low temperature solutions.


\begin{thebibliography}{99}


\bibitem{BCS}
J. Bardeen, L. N. Cooper, and J. R. Schrieffer,
``Theory of Superconductivity,''
Phys. Rev. 108, 1175 (1957)

\bibitem{Polchinski:1992ed}
  J.~Polchinski,
  ``Effective Field Theory And The Fermi Surface,''
  arXiv:hep-th/9210046.

\bibitem{hussey1}
R~A.~Cooper, Y.~Wang, B.~Vignolle, O.~J.~Lipscombe, S.~M.~Hayden, Y.~Tanabe, T.~Adachi,
Y.~Koike, M.~Nohara, H.~Takagi, C.~Proust and N.~E.~Hussey,
``Anomalous criticality in the electrical resistivity of La$_{2-x}$Sr$_x$CuO$_4$'',
Science, {\bf 323}, 603 (2009).

\bibitem{hussey2}
N.~E.~Hussey,
``Phenomenology of the normal state in-plane transport properties of high-T$_c$ cuprates,''
J. Phys.:Condens. Matter {\bf 20}, 123201 (2008)
[arXiv:0804.2984 [cond-mat.supr-con]].

\bibitem{Lee:2008xf}
  S.~S.~Lee,
  ``A Non-Fermi Liquid from a Charged Black Hole: A Critical Fermi Ball,''
  Phys.\ Rev.\  D {\bf 79}, 086006 (2009)
  [arXiv:0809.3402 [hep-th]].

\bibitem{Liu:2009dm}
  H.~Liu, J.~McGreevy and D.~Vegh,
  ``Non-Fermi liquids from holography,''
  arXiv:0903.2477 [hep-th].

\bibitem{Cubrovic:2009ye}
  M.~Cubrovic, J.~Zaanen and K.~Schalm,
  ``String Theory, Quantum Phase Transitions and the Emergent Fermi-Liquid,''
   Science {\bf 325}, 439 (2009)
  [arXiv:0904.1993 [hep-th]].

\bibitem{Faulkner:2009wj}
  T.~Faulkner, H.~Liu, J.~McGreevy and D.~Vegh,
  ``Emergent quantum criticality, Fermi surfaces, and AdS2,''
  arXiv:0907.2694 [hep-th].

\bibitem{Hartnoll:2009ns}
  S.~A.~Hartnoll, J.~Polchinski, E.~Silverstein and D.~Tong,
  ``Towards strange metallic holography,''
  arXiv:0912.1061 [hep-th].

\bibitem{Hartman:2009qu}
  T.~Hartman, W.~Song and A.~Strominger,
  ``The Kerr-Fermi Sea,''
  arXiv:0912.4265 [hep-th].

\bibitem{Faulkner:2010tq}
  T.~Faulkner and J.~Polchinski,
  ``Semi-Holographic Fermi Liquids,''
  arXiv:1001.5049 [hep-th].

\bibitem{Maldacena:1997re}
J.~M.~Maldacena, ``The large N limit of superconformal field
theories and supergravity,'' Adv.\ Theor.\ Math.\ Phys.\ {\bf 2}
(1998) 231 [Int.\ J.\ Theor.\ Phys.\ {\bf 38} (1999) 1113]
[arXiv:hep-th/9711200].

\bibitem{Hartnoll:2009sz}
  S.~A.~Hartnoll,
  ``Lectures on holographic methods for condensed matter physics,''
  arXiv:0903.3246 [hep-th].

\bibitem{Herzog:2009xv}
  C.~P.~Herzog,
  ``Lectures on Holographic Superfluidity and Superconductivity,''
  J.\ Phys.\ A  {\bf 42}, 343001 (2009)
  [arXiv:0904.1975 [hep-th]].

\bibitem{McGreevy:2009xe}
  J.~McGreevy,
  ``Holographic duality with a view toward many-body physics,''
  arXiv:0909.0518 [hep-th].

\bibitem{Hartnoll:2009qx}
  S.~A.~Hartnoll,
  ``Quantum Critical Dynamics from Black Holes,''
  arXiv:0909.3553 [cond-mat.str-el].

  \bibitem{ARPES}
A.~Damascelli, Z.~Hussain and Z-X.~Shen,
``Angle-resolved photoemission studies of the cuprate superconductors,''
Rev. Mod. Phys. {\bf 75}, 473 (2003).

\bibitem{Denef:2009yy}
  F.~Denef, S.~A.~Hartnoll and S.~Sachdev,
  ``Quantum oscillations and black hole ringing,''
  arXiv:0908.1788 [hep-th].

\bibitem{Hartnoll:2009kk}
  S.~A.~Hartnoll and D.~M.~Hofman,
  ``Generalized Lifshitz-Kosevich scaling at quantum criticality from the
  holographic correspondence,''
  arXiv:0912.0008 [cond-mat.str-el].

\bibitem{Gubser:2008px}
  S.~S.~Gubser,
  ``Breaking an Abelian gauge symmetry near a black hole horizon,''
  Phys.\ Rev.\  D {\bf 78}, 065034 (2008)
  [arXiv:0801.2977 [hep-th]].

\bibitem{Hartnoll:2008vx}
  S.~A.~Hartnoll, C.~P.~Herzog and G.~T.~Horowitz,
  ``Building a Holographic Superconductor,''
  Phys.\ Rev.\ Lett.\  {\bf 101}, 031601 (2008)
  [arXiv:0803.3295 [hep-th]].

\bibitem{Hartnoll:2008kx}
  S.~A.~Hartnoll, C.~P.~Herzog and G.~T.~Horowitz,
  ``Holographic Superconductors,''
  JHEP {\bf 0812}, 015 (2008)
  [arXiv:0810.1563 [hep-th]].

\bibitem{Denef:2009tp}
  F.~Denef and S.~A.~Hartnoll,
  ``Landscape of superconducting membranes,''
  Phys.\ Rev.\  D {\bf 79}, 126008 (2009)
  [arXiv:0901.1160 [hep-th]].

\bibitem{Faulkner:2009am}
  T.~Faulkner, G.~T.~Horowitz, J.~McGreevy, M.~M.~Roberts and D.~Vegh,
  ``Photoemission `experiments' on holographic superconductors,''
  arXiv:0911.3402 [hep-th].

\bibitem{Alford:2007xm}
  M.~G.~Alford, A.~Schmitt, K.~Rajagopal and T.~Schafer,
  ``Color superconductivity in dense quark matter,''
  Rev.\ Mod.\ Phys.\  {\bf 80}, 1455 (2008)
  [arXiv:0709.4635 [hep-ph]].

\bibitem{Bertrand}
D. Bertrand, ``A Relativistic BCS Theory of Superconductivity,''
Ph.D. thesis, Catholic University of Louvain (Louvain-la-Neuve, Belgium, July 2005), available at http://dial.academielouvain.be:8080/vital/access/manager/Repository/boreal:5380 \ .

\bibitem{BertrandGovaerts}
J. Govaerts and D. Bertrand,
``Superconductivity and Electric Fields: A Relativistic Extension of BCS Superconductivity,''
arXiv:cond-mat/0608084 (2006).

\bibitem{Denef:2009kn}
  F.~Denef, S.~A.~Hartnoll and S.~Sachdev,
  ``Black hole determinants and quasinormal modes,''
  arXiv:0908.2657 [hep-th].

\bibitem{Iqbal:2009fd}
  N.~Iqbal and H.~Liu,
  ``Real-time response in AdS/CFT with application to spinors,''
  Fortsch.\ Phys.\  {\bf 57}, 367 (2009)
  [arXiv:0903.2596 [hep-th]].

\bibitem{Porrati:2009dy}
  M.~Porrati and L.~Girardello,
  ``The Three Dimensional Dual of 4D Chirality,''
  JHEP {\bf 0911}, 114 (2009)
  [arXiv:0908.3487 [hep-th]].

\bibitem{Rattazzi:2009ux}
  R.~Rattazzi and M.~Redi,
  ``Gauge Boson Mass Generation in AdS4,''
  JHEP {\bf 0912}, 025 (2009)
  [arXiv:0908.4150 [hep-th]].

\bibitem{Son:2002sd}
  D.~T.~Son and A.~O.~Starinets,
  ``Minkowski-space correlators in AdS/CFT correspondence: Recipe and
  applications,''
  JHEP {\bf 0209}, 042 (2002)
  [arXiv:hep-th/0205051].

\bibitem{Ching:1995tj}
  E.~S.~C.~Ching, P.~T.~Leung, W.~M.~Suen and K.~Young,
  ``Wave propagation in gravitational systems: Late time behavior,''
  Phys.\ Rev.\  D {\bf 52}, 2118 (1995)
  [arXiv:gr-qc/9507035].

\bibitem{hertz}
  J.~A.~Hertz,
  ``Quantum critical phenomena,''
  Phys. Rev. B {\bf 14} (1976) 1165.

\bibitem{longrange1}
N.~E.~Bonesteel, I.~A.~McDonald and C.~Nayak,
``Gauge fields and pairing in double-layer composite fermion
metals,'' Phys. Rev. Lett. {\bf 77}, 3009 (1996).

\bibitem{longrange2}
J.-H.~She and J.~Zaanen,
``BCS superconductivity in quantum critical metals,''
Phys. Rev. {\bf B80}, 184518 (2009).

\bibitem{longrange3}
D.~T.~Son,
``Superconductivity by long-range color magnetic interaction in high-density quark matter,''
Phys. Rev. {\bf D59}, 094019 (1999).

\bibitem{Mateos:2007yp}
  D.~Mateos and L.~Patino,
  ``Bright branes for strongly coupled plasmas,''
  JHEP {\bf 0711}, 025 (2007)
  [arXiv:0709.2168 [hep-th]].

\bibitem{Rajagopal:2000wf}
  K.~Rajagopal and F.~Wilczek,
  ``The condensed matter physics of QCD,''
  arXiv:hep-ph/0011333.


\end{thebibliography}
\end{document}